\documentclass[aps,pre,twocolumn,float,showpacs]{revtex4}
\usepackage{amsmath,bm,epsfig}

\let \nn  \nonumber

\let\*\cdot
\def\<{\left\langle} \def\>{\right\rangle} \def\({\left(} \def\){\right)}
 \let\~\widetilde \let\^\widehat 

\def\be{\begin{equation}}\def\ee{\end{equation}}
\def\bea{\begin{eqnarray}}\def\eea{\end{eqnarray}}
\def\bse{\begin{subequations}}\def\ese{\end{subequations}}
\newcommand{\BE}[1]{\begin{equation}\label{#1}}
\newcommand{\BEA}[1]{\begin{eqnarray}\label{#1}}
\newcommand{\BSE}[1]{\begin{subequations}\label{#1}}

\let \nn  \nonumber
\usepackage{pstricks}
\usepackage{pst-node}
\usepackage[ansinew]{inputenc}
\usepackage{amssymb,amsmath}

\def\BSE{\begin{subequations}}\def\ESE{\end{subequations}}
\let \= \equiv

\def\a{\alpha}
\def\b{\beta}

\def\o{\omega}

\def\be{\begin{equation}}       \def\ba{\begin{array}}

\def\ee{\end{equation}}         \def\ea{\end{array}}

\def\bea {\begin{eqnarray}}      \def\eea {\end{eqnarray}}

\def\bean{\begin{eqnarray*}}    \def\eean{\end{eqnarray*}}

           \def\ph{\varphi}

\def\const {\mathop{\rm const}\nolimits}

\def\<{\langle} \def\({\left(}  \def\>{\rangle} \def\){\right)}

\begin{document}

\title{Resonance clustering in
wave turbulent regimes:\\ Integrable dynamics}
\author{Elena Kartashova$^{\dag}$, Miguel D. Bustamante}
 \email{lena@risc.uni-linz.ac.at, miguel.bustamante@ucd.ie}
\affiliation{$^{\dag}$RISC, J.Kepler University, Linz 4040, Austria \\
$^*$School of Mathematical Sciences, University College Dublin, Belfield, Dublin 4, Ireland
}

\begin{abstract}
Two fundamental facts of the modern wave turbulence theory are
1) existence of power energy spectra in $k$-space, and 2) existence of ``gaps" in this spectra corresponding to the resonance clustering. Accordingly, three wave turbulent regimes are singled out: \emph{kinetic}, described by wave kinetic equations and power energy spectra; \emph{discrete}, characterized by resonance clustering; and \emph{mesoscopic},
where both types of wave field time evolution coexist. In this paper we study integrable dynamics of resonance clusters appearing in discrete and mesoscopic wave turbulent regimes.  Using
a novel method based on the notion of dynamical invariant we establish that some of the frequently met clusters are
integrable in quadratures for arbitrary initial conditions and some others -- only for particular initial conditions. We
also identify chaotic behaviour in some cases. Physical implications of the results obtained are discussed.
\end{abstract}



\maketitle \tableofcontents

\section{Introduction}

The broad structure of modern nonlinear science born at the edge of physics and mathematics includes an enormous number of
applications in cosmology, biochemistry, electronics, optics, hydrodynamics, economics, neuroscience, etc.  The emergence
of  nonlinear science itself as a collective interdisciplinary activity is due to the awareness that its dynamic concepts
first observed and understood in one field (for example, population biology, flame-front propagation, non-linear optics or
planetary motion) could be useful in others (such as in chemical dynamics, neuroscience, plasma confinement or weather
prediction).  The theory of integrable Hamiltonian systems, a generalization of the classical theory of differential
equations, is the nucleus of the whole nonlinear science. Various classifications of integrable systems are presently known
which turned out to be quite useful for physical applications. Classifications are known based on the various intrinsic
properties of integrable systems \cite{Olv93}: symmetries, conservation laws, Lax-pairs, etc.  In \cite{BF04} the general
classification of integrable Hamiltonian systems is presented based on the form of their topological invariants. The
usefulness of this classification is demonstrated in several problems on solid mechanics. In particular, it is proven that
two famous problems -- the Euler case in rigid body dynamics and the Jacobi problem of geodesics on the ellipsoid-- are
orbitally equivalent. In \cite{DZh01} the idea of classification is presented based on normal forms of a certain class of
bi-hamiltonian PDEs. Miscellaneous hierarchies of integrable PDEs are presented in \cite{MSS91}.

\begin{figure*}
\begin{center}\includegraphics[width=14cm,height=5.5cm]{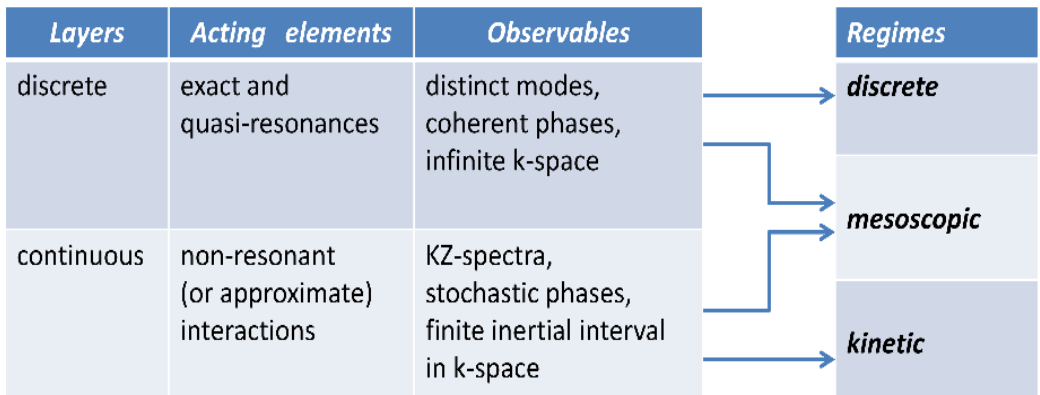}
\end{center}
\caption{\label{TTT} Color online. Schematic representation of wave turbulent regimes.}
\end{figure*}

The list can be prolonged further but the main point for us presently is the following: the notion of integrability itself
is ambitious! There are many quite different definitions of integrability, for instance integrability in terms of
elementary functions (equation $\ddot{y}=-y$ has the explicit solution $y=a \sin{(x+b)}$ ); integrability \emph{modulo}
class of functions (equation $\ddot{y}=f(y)$ has general solutions in terms of elliptic functions), etc. An example of less
obvious definition of integrability is C-integrability, first introduced in \cite{C91}: integrability \emph{modulo} change
of variables, meaning that a \emph{nonlinear} equation is called C-integrable if it can be turned into a \emph{linear}
equation by an appropriate invertible change of variables. For instance, Thomas equation $\psi_{xy}+\a \psi_x + \b \psi_y +
\psi_x \psi_y=0$ is C-integrable. Profound discussion on the subject can be found in \cite{KSh05}. In the present paper,
integrability is interpreted in terms of the existence of a number of independent dynamical invariants of the system; for
each in-this-sense-integrable system, solutions are then written out \emph{in quadratures}.

The dynamical systems we are interested in, describe nonlinear resonance clusters appearing in evolutionary dispersive PDEs
in two space variables. Nonlinear resonances are ubiquitous in physics. They appear in a great amount of typical mechanical
systems \cite{Mech1,Mech2}, in engineering \cite{Turbine,Cretin,Laser,Helium2}, astronomy
\cite{Astr}, biology \cite{Biology}, etc. etc. Euler equations, regarded with various boundary conditions and specific values of some parameters, describe an enormous
number of nonlinear dispersive wave systems (capillary waves, surface water waves, atmospheric planetary waves, drift waves
in plasma, etc.) all possessing nonlinear resonances.

The classical approach of statistical wave turbulence theory in
a nonlinear wave system assumes weak nonlinearity, randomness of phases, infinite-box limit, and existence of an inertial
interval in wavenumber space $(k_0, k_1)$, where energy input and dissipation are balanced. Under these assumptions, the
wave system is energy conserving, and wave kinetic equations describing the wave spectrum have stationary solutions in the
form of Kolmogorov-Zakharov (KZ) energy power spectra $k^{-\a}, \, \a>0, $ (\cite{Ph60,zak2,zakh92}, etc.).

As it was first established in the frame of the model of laminated turbulence,
\cite{K06-1},  KZ-spectra\index{KZ-spectra!gaps} have ``gaps" formed by exact and quasi-resonances (that is, resonances with small enough resonance broadening). This yields two distinct layers of turbulence in an arbitrary nonlinear wave system -- continuous and discrete -- and their interplay generates three possible wave turbulent regimes: \emph{kinetic}, \emph{discrete} and \emph{mesoscopic} as it is shown in Fig. \ref{TTT} (see \cite{CKW09} for more discussion).

The existence of mesoscopic regime has been first confirmed in numerical simulations with dynamical equations for surface gravity waves in \cite{zak4}, while discrete regime has been first described in \cite{K09b}. These theoretical findings are confirmed by numerous laboratory experiments.
For instance, in the experiments with gravity surface wave turbulence in a laboratory flume, \cite{de}, only a discrete regime has been identified while in \cite{WBP96} coexistence of both types of time evolution has been established.
Taking into account additional physical parameters in a wave system  transition from kinetic to mesoscopic regime can be observed as it was demonstrated in \cite{CK09} for capillary water waves, with and without rotation.

From a mathematical point of view, the very special role of resonant solutions has been first demonstrated by Poincar\'e
 who proved, using Calogero's terminology, that a nonlinear ODE is C-integrable if it has no resonance
solutions (see \cite{Arn3} and refs. therein). This statement allows the following
 Hamiltonian formulation \cite{zakh92}:%
 \bea i\,\dot a_{\bf k} &=&
\partial {\cal H}/\partial a_{\bf k}^*,
\label{HamiltonianEquationOfMotion} \eea
 \noindent
  where $a_{\bf k}$ is the amplitude of the Fourier mode corresponding to the
  wavevector ${\bf k}$ and the Hamiltonian ${\cal H}$ is represented
as an expansion in powers ${\mathcal{H}}_j$ which are proportional to the product of $j$ amplitudes $a_{\bf k}$. Then the
cubic Hamiltonian
has the form%
\bea {\cal H}_3= \sum_{{\bf k_1}, {\bf k_2},{\bf k_3}} V^3_{12}
a_{1}^* a_2 a_3\delta^1_{23}+\mbox{ complex conj.}, \nonumber
\label{QubicHamiltonian} \eea %
 \noindent
  where for brevity we
introduced the notation $ a_j \equiv a_{{\bf k}_j}$ and $
\delta^1_{23} \equiv \delta({\bf k_3} - {\bf k_1} - {\bf k_2}) $ is
the Kronecker symbol. If ${\cal H}_3 \ne 0$, three-wave resonant
processes are dominant. These satisfy the resonance conditions: %
 \bea
\label{res}
\begin{cases}
\omega ({\bf k}_1) + \omega ({\bf k}_2)- \omega ({\bf k}_{3}) = 0\\
{\bf k}_1 + {\bf k}_2 -{\bf k}_{3} = 0,
\end{cases}
\eea %
 \noindent  where $\omega({ \bf k})$ is a
dispersion relation for the linear wave frequency. Further on, the notation $\o_k$ is used for $\omega({ \bf k}).$ The
corresponding dynamical system has a general form%
 \bea \label{EquationOfMotion3Wd}
i\dot B_{\bf k} =  \sum_{{\bf k_1}, {\bf k_2}} \big( V^{\bf k}_{12}
B_{1} B_2 \delta^{\bf k}_{12}
+ 2 V^{1 \, *}_{{\bf k}2}  B_1 B_{2}^*  \delta^1_{{\bf k}3} )\eea
(notations  $B_j$ are used further on for the resonant modes). If ${\cal H}_3 = 0,$ four-wave resonances have to be studied, and so on. To confirm that ${\cal H}_3 \ne 0$ and three-wave
resonances are dominant, one has to find solutions of (\ref{res}) and check that $V^{\bf 3}_{12} \neq 0$ at least at some
resonant triads. Afterwards the corresponding dynamical system has to be studied.

It has been first proven in \cite{AMS}
that for a big class of physically relevant dispersion functions $\o$, the set of all wavevectors satisfying (\ref{res}) can be divided into
non-intersecting classes and solutions of (\ref{res}) can be looked for in each class separately. The method of $q$-class decomposition first introduced in
\cite{K06-3} has been developed specially for solving systems of the form (\ref{res}) in integers;  details of its
implementation for various rational and irrational dispersion functions are given in \cite{KK06-1}--\cite{KK07}. General description of the $q$-class method and corresponding programming codes are given in \cite{K09-2}, in Ch.3 and Appendix correspondingly.

An immediate consequence of the $q$-class method is that dynamical system (\ref{EquationOfMotion3Wd}) can be reduced
to \emph{a few dynamical systems} of smaller order, and each of these smaller dynamical systems can be investigated
independently from all others. In \cite{KM07}, construction of a set of reduced dynamical systems corresponding to the
solutions of (\ref{res}) and the systems themselves are given explicitly (as an example, resonances of oceanic planetary
waves were considered in the spectral domain $0 \le m,n \le 50$). The integrability of some resonance clusters has been
studied in \cite{BK09_1}.

The main goal of the present paper is to study systematically the integrable dynamics of the most frequently met resonance
clusters. We begin with a brief introduction of NR-diagrams (NR for nonlinear resonance) which give a handy graphical representation of a
generic resonance cluster and allow us to recover uniquely the dynamical system corresponding to each cluster \cite{K09-2}.

\section{\textbf{NR}-diagrams}

In systems with cubic Hamiltonian, a resonant triad is called  \emph{primary cluster} \cite{KK07} (a resonant quartet is a primary cluster in
a system with quadric Hamiltonian, and so on).  All other clusters (formed by a few primary clusters connected via one of a few joint modes) are called \emph{generic} clusters or simply
clusters. The dynamical system for a complex triad in the standard Manley--Rowe form (that is, with one interaction
coefficient $Z$) reads
\be \label{dyn3waves}  \dot{B}_1=   Z B_2^*B_3,\quad
\dot{B}_2= Z B_1^* B_3, \quad \dot{B}_3= -  Z B_1 B_2, %
\ee 
and is known to be integrable (e.g. \cite{book-triad}), with
two conservation laws in the Manley--Rowe form being%
 \be \nonumber \label{laws3waves}
 I_{23}=|B_2 |^2 + |B_3|^2,  \  I_{13}= |B_1 |^2 + |B_3|^2.
\ee
Due to the criterion of nonlinear instability for a triad \cite{H67}, the mode with maximal frequency, $\o_3$, is unstable
while the modes $\o_1$ and $\o_2$ are neutral. Originally, this fact has been deduced directly from the equations of motion, \cite{H67}, but it can easily be seen from the the form of
 Manley--Rowe constants, \cite{KL-08}.

This means that the form of dynamical systems and accordingly time evolution of the modes belonging to a generic cluster depends crucially on the fact whether joint modes within a cluster are stable or unstable. With the purpose to distinguish between these cases, the
notations A-mode
(active) and P-mode (passive) are introduced  for $\o_3$-mode and $\o_1$- and $\o_2$-modes respectively, \cite{KL-08}. This allows to describe all possible connection types within a generic cluster. For instance, 1-mode connection of two triads
can be of AA-, AP- and PP-type; 1-mode connection of three triads can be of AAA-, AAP-, APP-type and PPP-type, 2-mode
connection between two triads can be of AA-PP-, AP-AP-, AP-PP- and PP-PP-types, and so on.

\begin{figure}
\begin{center} \includegraphics[height=4cm]{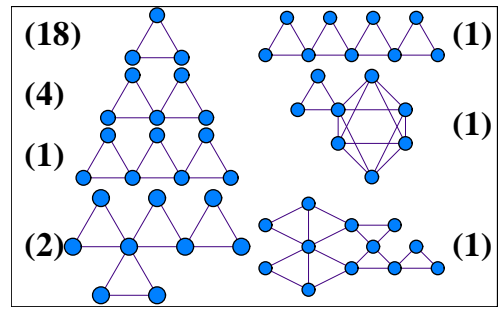}
\end{center}
\caption{\label{f:top} Topological structure of the cluster set for
the oceanic
planetary waves, $\o  \sim 1/\sqrt {m^2+n^2}$, in the domain $m,n \le
50$. 7
types of resonance clusters have been found, the number of the
clusters of each type is shown in parenthesis.}
\end{figure}

In the topological representation \cite{KM07} of the
solution set of (\ref{res}) this dynamical information has been kept unexplicit, as part of a programming code used to
construct dynamical system, while each triad within a cluster was shown as an unmarked triangle (see Fig.\ref{f:top}). In this representation each vertex shown as a circle denotes \emph{one resonant mode}. This representation has been slightly improved in  \cite{KL-08} where each $A$-mode has been marked by two arrow-edges coming from $A$-mode to both $P$-modes in each triad. However, for a larger clusters this representation becomes too nebulous.
\begin{figure}
\begin{center}\includegraphics[width=8cm,height=5.5cm]{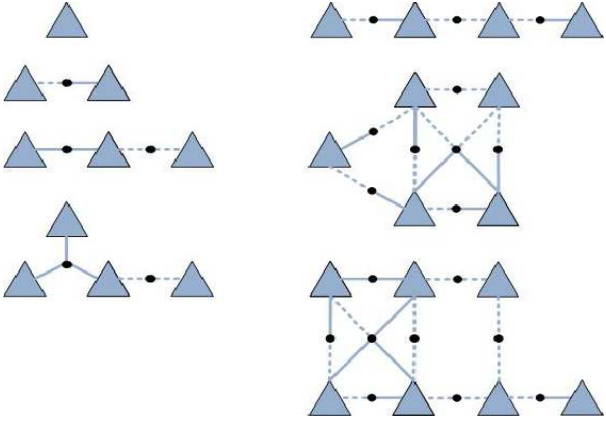}
\end{center}
\caption{\label{NB-diag} NR-diagrams for some resonance clusters shown in Fig.\ref{f:top}.}
\end{figure}

More compact graphical representation of a resonance cluster is given by its NR-diagram
first introduced in \cite{K09b}, both for three- and four-wave resonance systems.  In a NR-diagram each vertex represents not a resonant mode but a \emph{primary cluster}, that is, a triad and a quartet in a three- and four-wave system correspondingly.

A
NR-diagram in systems with cubic Hamiltonian consists of following building elements  -- a triangle and two types of half-edges, bold
for A-mode and dotted for P-mode. It can be proven (\cite{K09-2}, Ch.3) that in this case the form of NR-diagram defines \emph{uniquely} corresponding dynamical system. Examples of NR-diagrams for some resonance clusters shown in Fig.\ref{f:top} are displayed in Fig.\ref{NB-diag}.

 Below,
dynamical systems are given for three generic clusters shown in Fig.\ref{f:top}, on the left (dynamical systems for the
other clusters are omitted for sake of space):

1. Cluster consisting of two triads $a$ and $b$, whose connecting mode is active in one triad and passive in the other
triad, say $B_{3a}=B_{1b}.$ In other words, a cluster with one AP-connection. It is called AP-butterfly \cite{KL-08} and
its dynamical system is
 \bea
 \begin{cases}\label{AP}
 \dot{B}_{1a}=  Z_a B_{2a}^*B_{3a}\,, \quad  \dot{B}_{3b}=  - Z_{b} B_{3a} B_{2b}\,, \\
\dot{B}_{2a}=  Z_a B_{1a}^* B_{3a}\,,\quad    \dot{B}_{2b}=  Z_b B_{3a}^* B_{3b}\,, \\
\dot{B}_{3a}=  - Z_{a} B_{1a} B_{2a}  + Z_b B_{2b}^* B_{3b},
 \end{cases}
\eea%

2. Cluster consisting of three triads $a$, $b$ and $c$, with one AA- and one PP-connections, say, $B_{3a}=B_{3b}$ and
$B_{1b}=B_{1c}.$  The dynamical system reads
 \bea
 \begin{cases}\label{AA-PP}
\dot{B}_{1a}=   Z_a B_{2a}^*B_{3a},\quad \dot{B}_{2a}= Z_a B_{1a}^* B_{3a}, \\
\dot{B}_{3a}= -  Z_a B_{1a} B_{2a} -  Z_b B_{1b} B_{2b}\\
\dot{B}_{1b}=   Z_b B_{2b}^*B_{3a} + Z B_{2c}^*B_{3c},\\
\dot{B}_{2b}= Z_b B_{1b}^* B_{3a}, \ \
\dot{B}_{2c}= Z_c B_{1b}^* B_{3c}, \\
\dot{B}_{3c}= -  Z_c B_{1b} B_{2c}.
 \end{cases}
\eea

3. Cluster consisting of four triads $a$, $b$,  $c$ and  $d$, with two AA- and one PP-connections, say,
$B_{3a}=B_{3b}=B_{3c}$ and $B_{1c}=B_{1d}.$ The corresponding dynamical system is of the form
 \bea
  \begin{cases}\label{2AA-PP}
\dot{B}_{1a}=   Z_a B_{2a}^*B_{3a},\ \
\dot{B}_{2a}= Z_a B_{1a}^* B_{3a}, \\
\dot{B}_{3a}= -  Z_a B_{1a} B_{2a} -  Z_c B_{1c} B_{2c} -  Z_d B_{1d} B_{2d}, \\
\dot{B}_{1b}=   Z_b B_{2b}^*B_{3a},\ \
\dot{B}_{2b}= Z_b B_{1b}^* B_{3a}, \\
\dot{B}_{3b}= -  Z_b B_{1b} B_{2b}\\
\dot{B}_{1c}=   Z_c B_{2c}^*B_{3a} + Z_d B_{2d}^*B_{3d},\\
\dot{B}_{2c}= Z_c B_{1c}^* B_{3a}, \ \
\dot{B}_{2d}= Z_d B_{1c}^* B_{3d}.
 \end{cases}
\eea

The Manley--Rowe constants can be written out immediately for each of these systems, being combinations of corresponding
constants for each triad. For instance, for (\ref{AP}) they have the form \bea
  \begin{cases}
I_{12b}=|B_{1b}|^2 - |B_{2b}|^2, \quad I_{23b}=|B_{2b}|^2 + |B_{3b}|^2,\\
I_{ab}= |B_{1b}|^2 + |B_{3a}|^2+|B_{3b}|^2.
 \end{cases}
\eea

NR-diagrams also give us immediate qualitative information about
 the energy percolation within bigger resonance clusters,
 for the case when two P-modes, forming a PP-connection, have small initial amplitudes compared to
 the amplitudes of A-modes in the connected triads. It follows then from the Hasselmann's criterion of
 instability \cite{H67} that a PP-connection can be regarded
as an obstacle for the energy percolation in both directions and this cluster can be then regarded practically as two
independent triads. Analogous considerations show that AA-connection allows energy percolation in two directions while
AP-connection - in one direction. In this sense one can define, for \emph{certain initial conditions}, PP-reductions of resonance
clusters, whereby a cluster is approximated by smaller clusters obtained by cutting off the PP-connections from the
original cluster.

While planning laboratory experiments, one has to be very careful with these theoretical findings. For instance, in the experiments reported in \cite{CHS96}, a chain-like cluster of three connected triads has been identified with one PP-connection. However, this connection could not be disregarded while  a very small ``parasite" frequency generated by electronic equipment was enough for initiating the energy exchange among the modes of all three triads (see \cite{K09-2}, Ch.4, for detailed explanations).

The main difference between NR-diagram and  statistical diagrams used in wave turbulence theory and originated from Feynman diagrams can be formulated as follows. Each statistical diagram corresponds to \emph{one term} in the asymptotic expansion and does not allow to compute the amplitudes of the scattering process. On the other hand, a NR-diagram describes \emph{completely} a resonance cluster and allows to write out explicit form of the dynamical system on the modes' amplitudes.

As it will be shown below, connection types within a cluster define indeed the integrability of the corresponding dynamical
systems. In order to demonstrate it we will use the notion of dynamical invariant first introduced in \cite{BK09_1} which is given in the next section and
illustrated by the example of harmonic oscillator.

\section{Dynamical invariants}
\subsection{Definition}

From here on, general notations and terminology will follow Olver's
book \cite{Olv93} and Einstein convention on repeated indices and
$f_{,i} \equiv \partial f/\partial x^i$. Consider a general
$N$-dimensional system of autonomous evolution equations of the
form:
 \be \label{m: evol}\frac{d{x}^i}{dt}(t) = \Delta^i(x^j(t)), \quad i=1, \ldots, N.
  \ee
 Any scalar function $f(x^i,t)$ that
 satisfies $$\frac{d}{dt}\left(f(x^i(t),t)\right) = \frac{\partial}{\partial t} f + \Delta^i f_{,i}  =
 0$$
is called {\it a conservation law} in \cite{Olv93}. It is easy to see that this definition gives us two types of
conservation laws: (i) those of the form $f(x^i)$ (no explicit time-dependence), and (ii) those of the form $f(x^i,t)$,
where the time dependence is explicit. The first type determines an invariant manifold for the dynamical system (\ref{m:
evol}) (\emph{time-independent} conservation law) and the second type constrains the time evolution of the system within the invariant manifold(s) (\emph{time-dependent} conservation law). To keep in mind
the difference between these two types of conservation laws, we call the first type just a conservation law (CL), and the
second type - a \emph{dynamical invariant}.

We are interested in determining the solution $x^i(t), \quad i=1, \ldots, N,$ of a given dynamical system of the form
(\ref{m: evol}). One possible way to do that is by finding $N$ functionally independent dynamical invariants for the system
(\ref{m: evol}). This is equivalent to finding $(N-1)$ functionally independent conservation laws and \emph{one} dynamical
invariant (the equivalence can be proven, for example, using the implicit function theorem).

As it was shown in \cite{BK09_2}, in some cases the knowledge of only $(N-2)$ functionally independent CLs is enough for
constructing explicitly: (i) a new CL functionally independent of the others, and (ii) a corresponding dynamical invariant,
 determining the solution $x^i(t), \quad i=1, \ldots, N.$ This follows from the Theorem on $(N-2)$--integrability
\cite{BK09_2}, whose formulation is given below for the readers' convenience.

{\textbf{Theorem on $(N-2)$--integrability.}}  \emph{Let us assume that the system (\ref{m: evol}) possesses a standard
Liouville volume density
$$\rho(x^i): \ ( \rho \Delta^i)_{,i} = 0,$$ and $(N-2)$ functionally
independent CLs, $H^1, \ldots, H^{N-2}$. Then a new CL in quadratures can be constructed, which is functionally
independent of the original ones, and therefore the system is integrable.}\\

\begin{figure*}
\begin{center}
\includegraphics[width=5cm,height=5cm,angle=270]{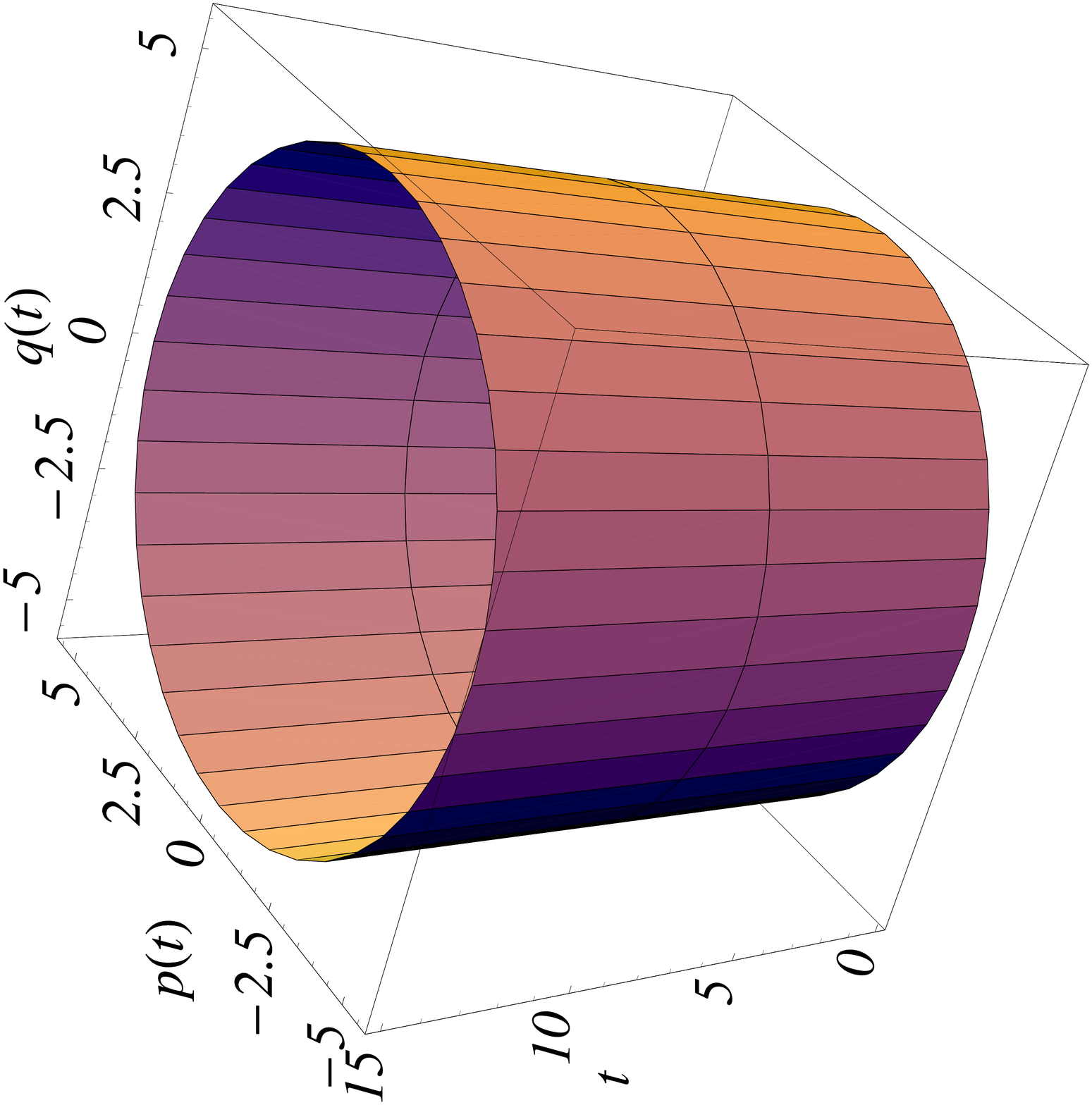}
\includegraphics[width=5cm,height=5cm,angle=270]{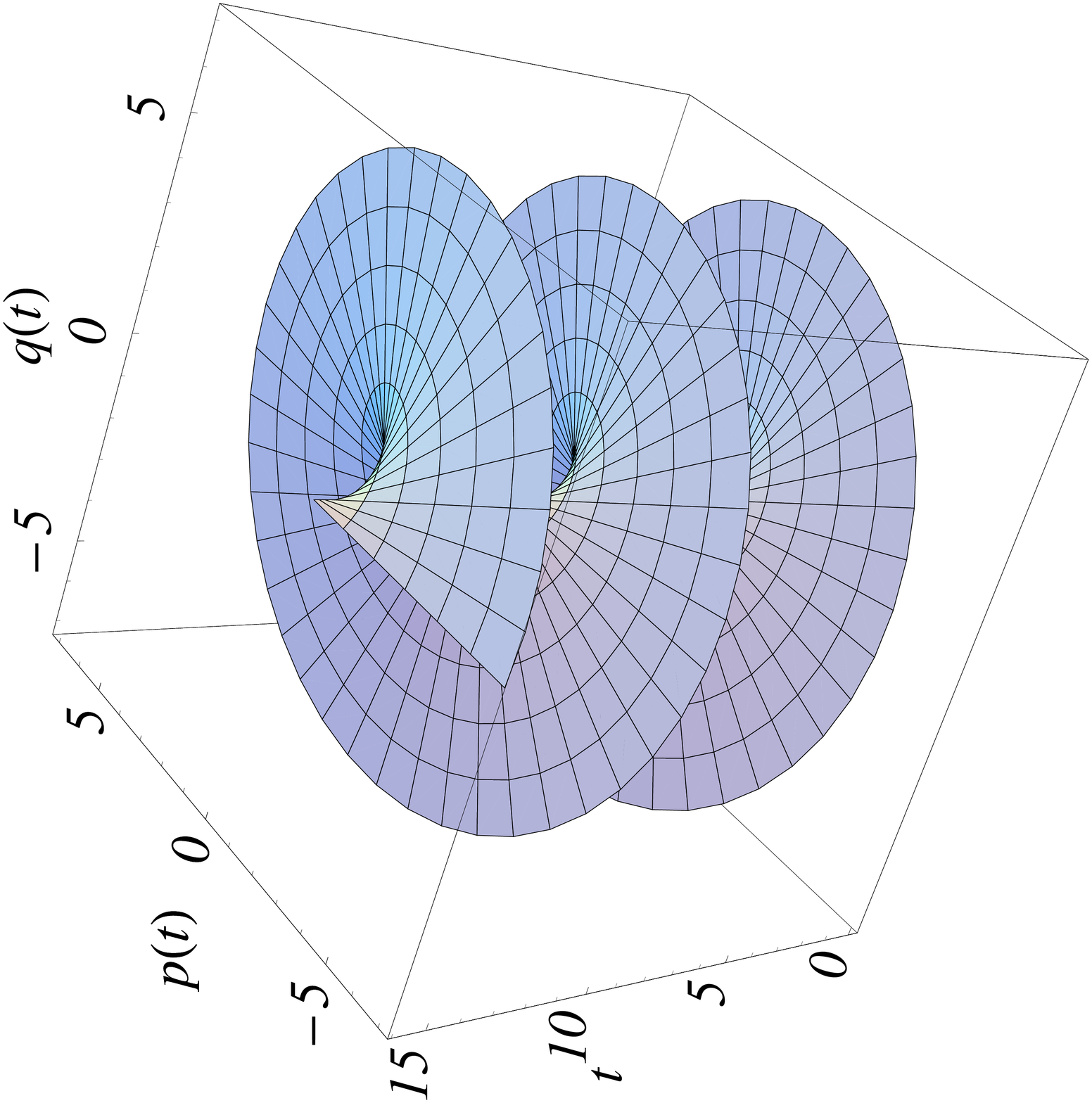}
\includegraphics[width=5cm,height=5cm,angle=270]{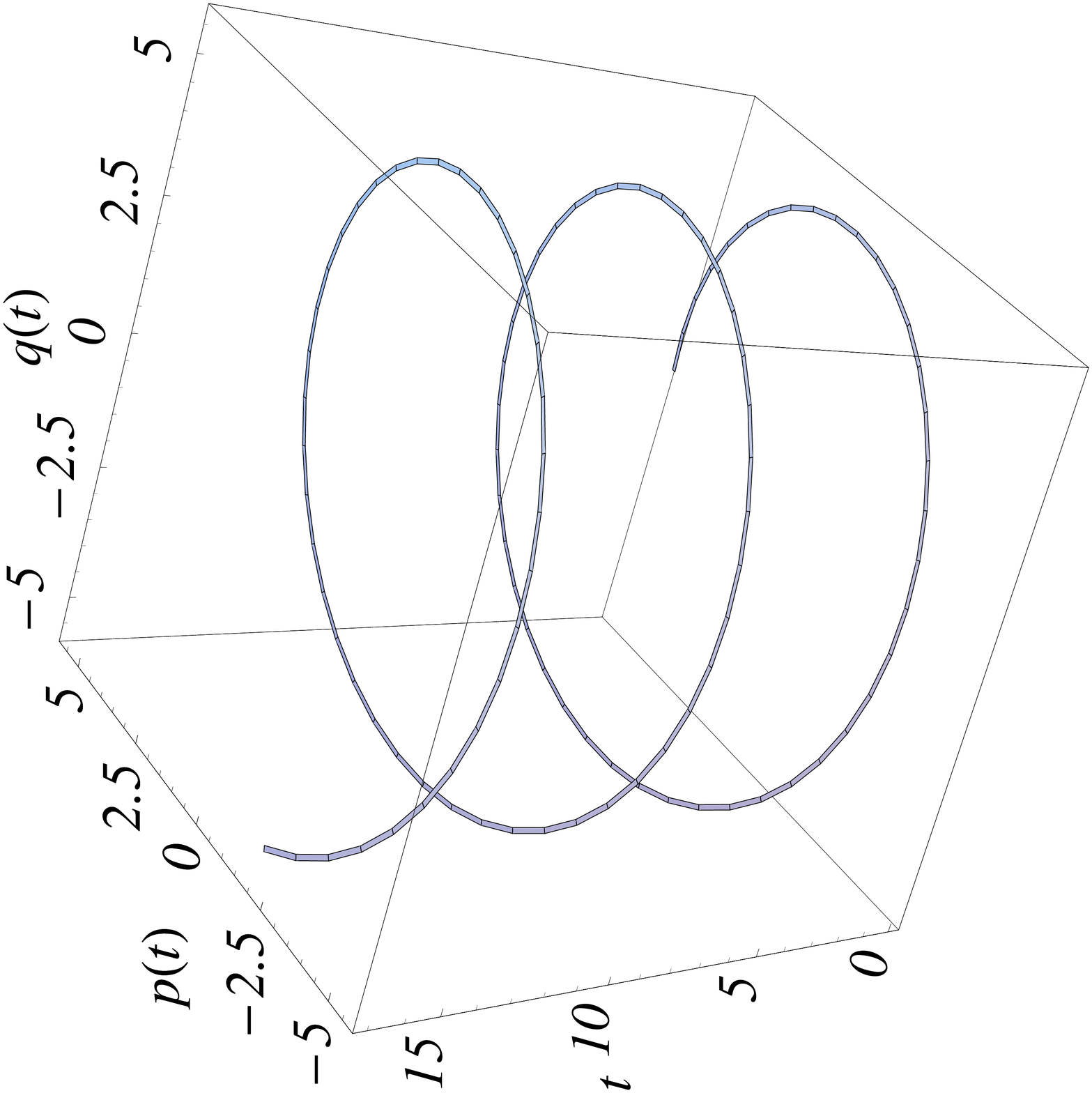}
\includegraphics[width=5cm,height=5cm,angle=270]{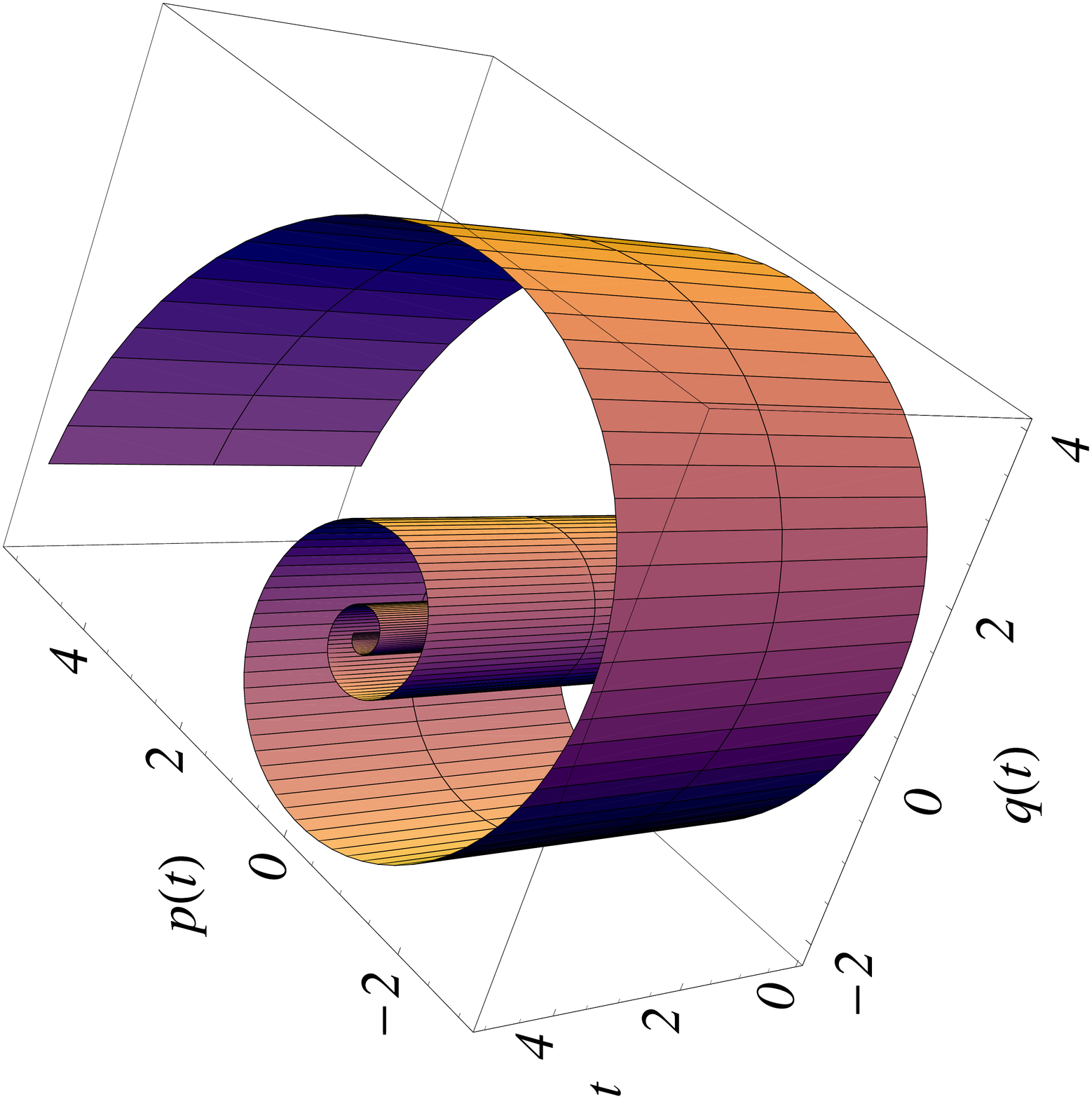}
\includegraphics[width=5cm,height=5cm,angle=270]{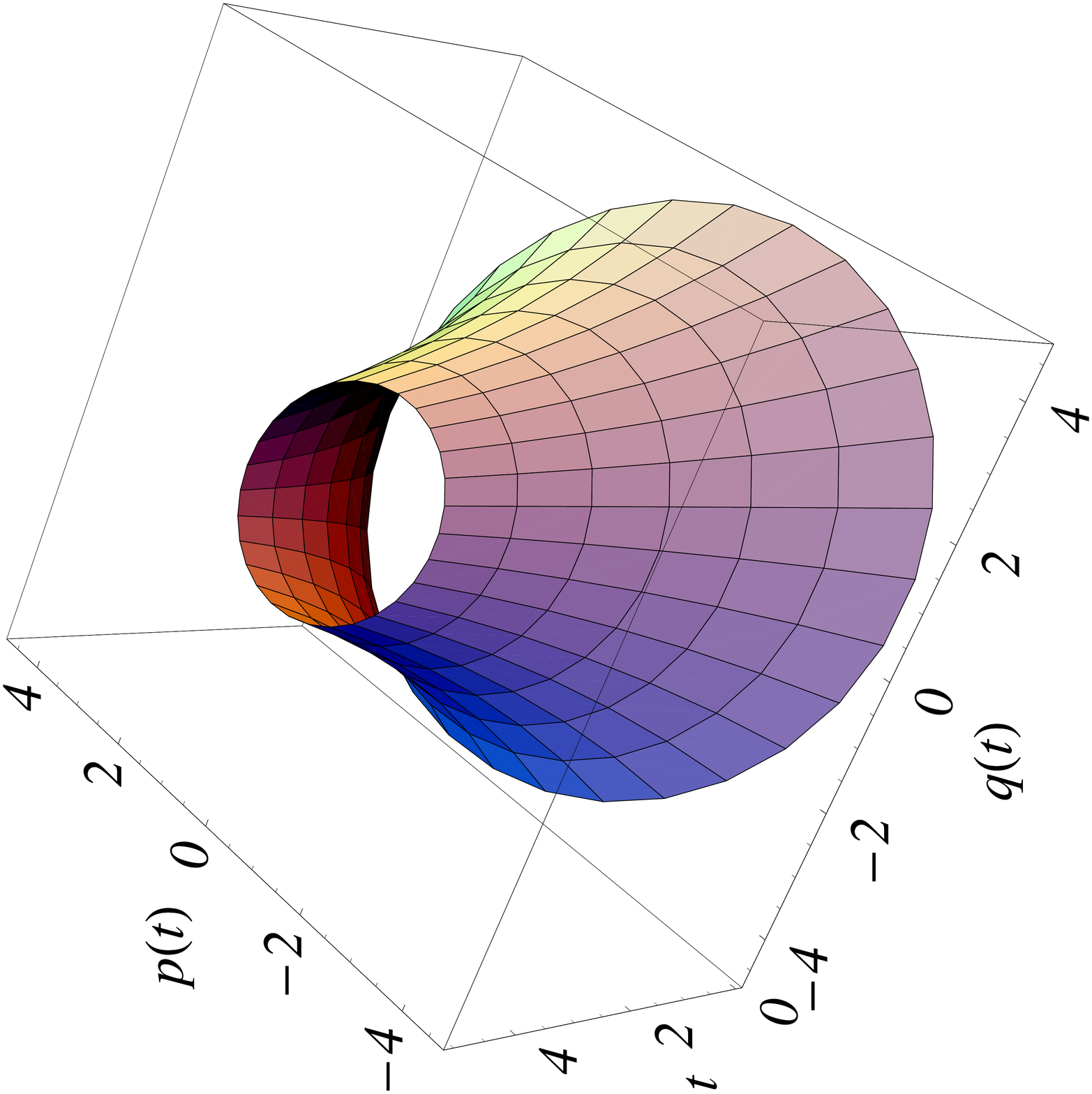}
\includegraphics[width=5cm,height=5cm,angle=270]{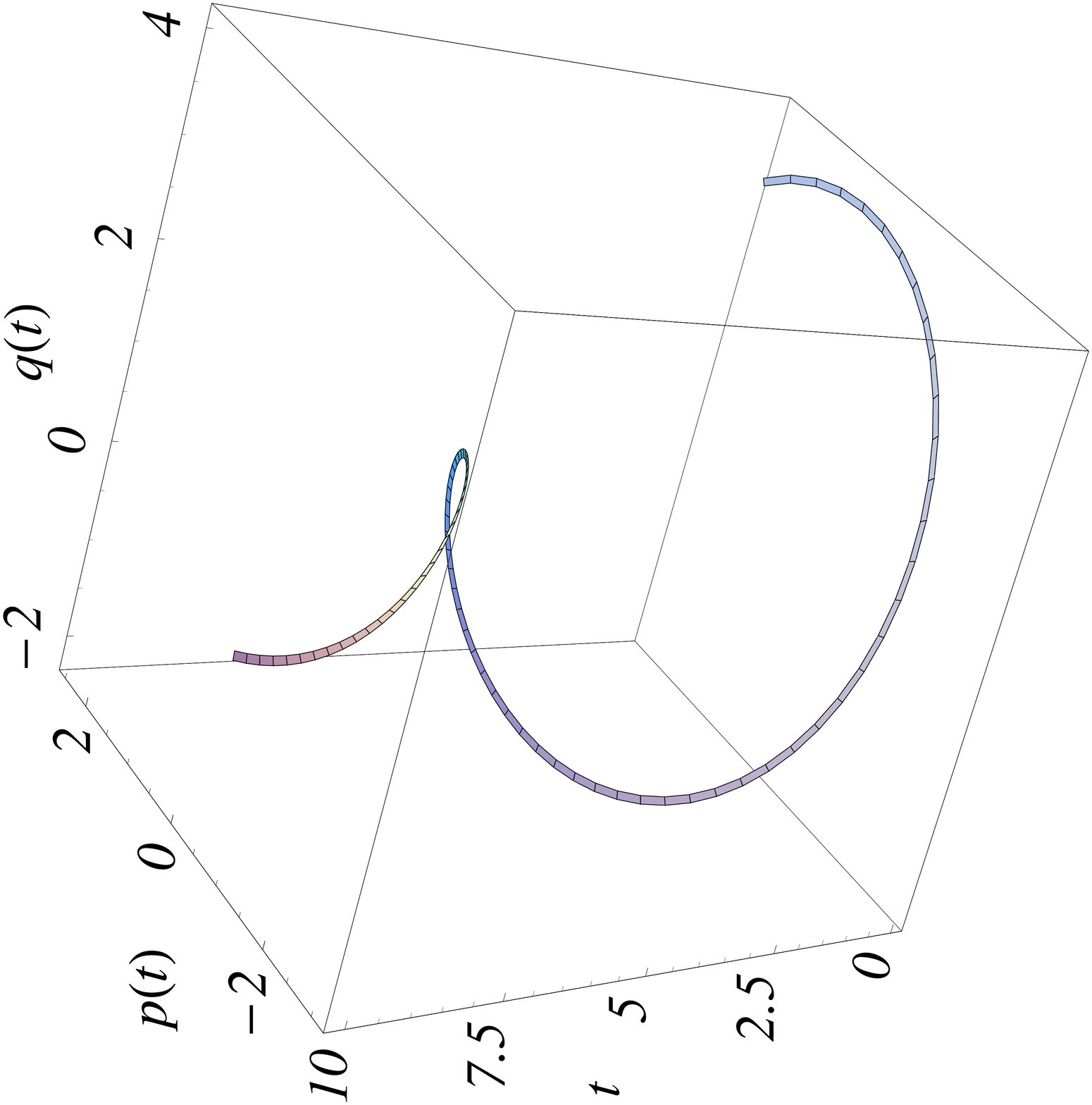}
\end{center}
\caption{\label{f:HO_parts} Color online. Upper panel: harmonic oscillator (case $\a=0$): left upper panel, a level surface of
conservation law
$E(q,p)=30$, Eq.(\ref{eq:E_HO}); middle upper panel, a level surface of dynamical invariant $T(q,p,t)=0$, Eq.(\ref{eq:Tnew_HO});
right upper
panel, solution trajectory $(q(t),p(t))$, Eq.(\ref{eq:sol_HO}), corresponding to $E=30,\,T=0$. Lower panel:
sub-critically damped harmonic
oscillator (case $0<\a<2$): left lower panel, a level surface of conservation law $C(q,p)=30$, Eq.(\ref{eq:Cnew_DHO});
middle upper panel, a
level surface of dynamical invariant $D(q,p,t)=20$, Eq.(\ref{eq:D_DHO}); right upper panel, solution trajectory $(q(t),p(t))$,
Eq.(\ref{eq:sol_DHO}), corresponding to $C=30,\,D=20$.}
\end{figure*}

\begin{figure*}
\begin{center}
\includegraphics[width=6cm,height=6cm,angle=270]{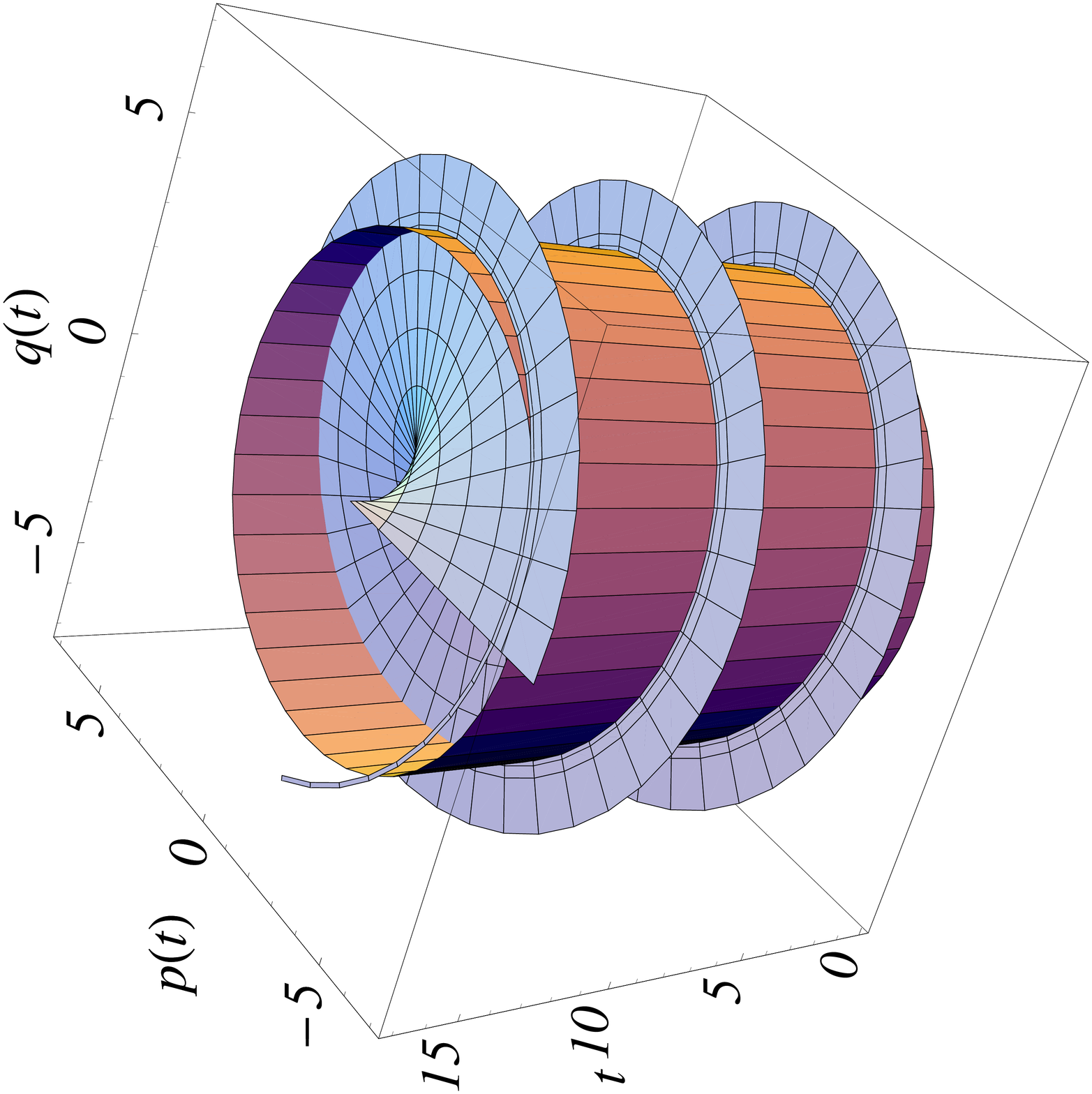}
\includegraphics[width=6cm,height=6cm,angle=270]{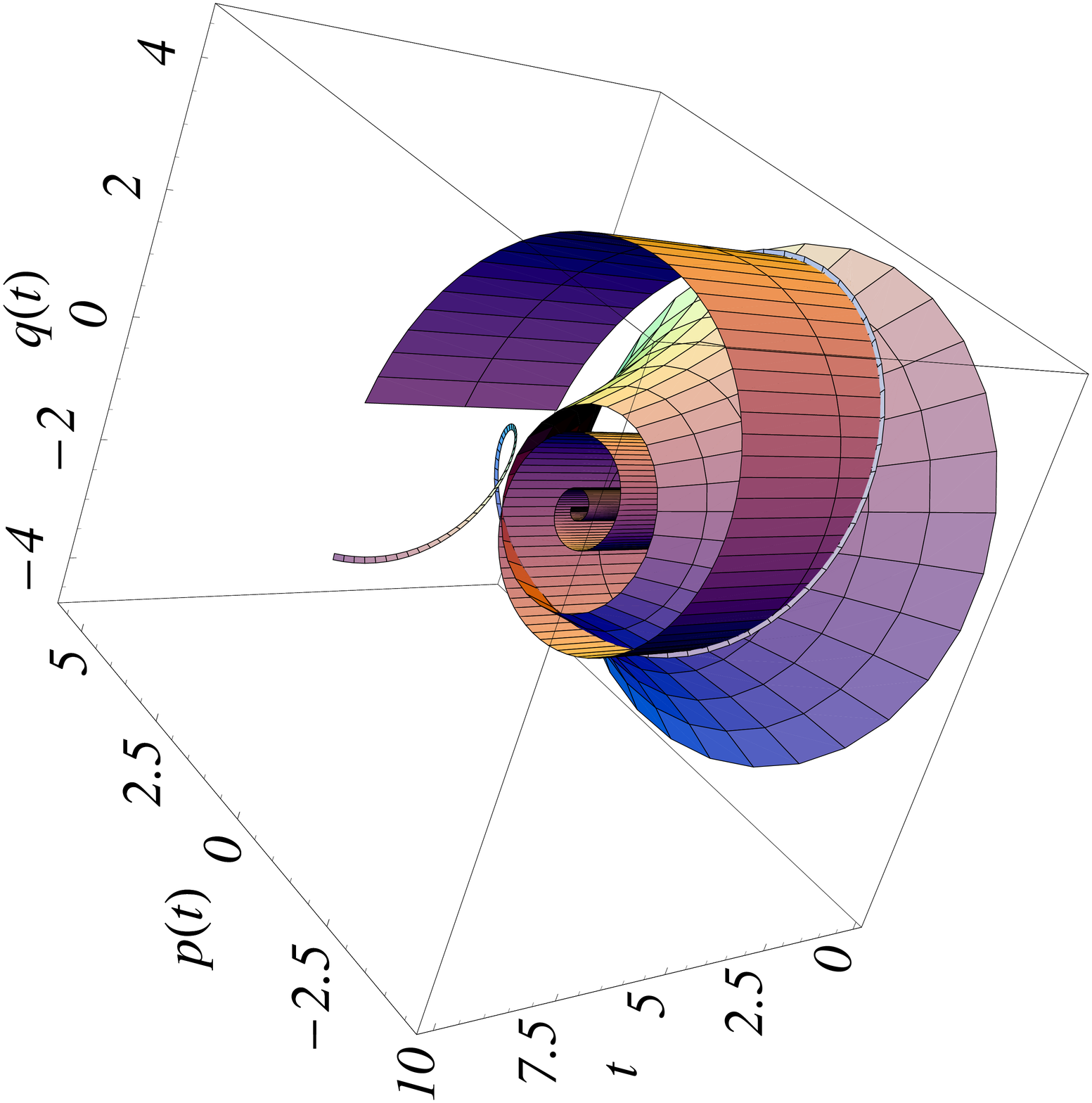}
\end{center}
\caption{\label{f:HO_all} Color online. Left panel: harmonic oscillator (case $\a=0$): combined plot of level
surface of conservation law
$E(q,p)$, level surface of dynamical invariant $T(q,p,t)$ and solution trajectory $(q(t),p(t))$. Notice the
general property that the
intersection of the level surfaces of $E$ and $T$ \emph{is} the solution trajectory. Right panel:
sub-critically damped harmonic oscillator (case
$0<\a<2$): combined plot of level surface of conservation law $C(q,p)$, level surface of dynamical
invariant $D(q,p,t)$ and solution trajectory
$(q(t),p(t))$. Notice the general property that the intersection of the level surfaces of $C$ and $D$ \emph{is}
the solution trajectory.}
\end{figure*}

\subsection{Example: damped harmonic oscillator}
\subsubsection{Dynamical invariants, CLs and solutions}

\emph{To illustrate the complementarity of conserved laws and
dynamical invariants}, we present an illustrative example from
mechanics for the case $N=2$. Consider the damped harmonic
oscillator. The equations of motion in non-dimensional form can be
written as:
\be \label{eq:DHO} \dot{q}=p\,, \quad \dot{p} = -q - \alpha p, \ee
where $\a \geq 0$ is the damping coefficient. This is a dynamical
system of the form (\ref{m: evol}) with $N=2$.

Now we want to fully determine the solution of the dynamical system
(\ref{eq:DHO}). For this we need to know both a CL and a dynamical
invariant. Indeed, let us consider separately the cases $\a=0$
(harmonic oscillator) and $0<\a<2$ (sub-critically damped harmonic
oscillator).

1. Case $\a=0$. We have the CL 
\be \label{eq:E_HO} E(q,p) = 1/2 \left(p^2 + q^2\right) \ee 
(energy) and the dynamical invariant
\be \label{eq:T_HO} T(q,p,t) = t - \arctan(q/p). \ee
Since
$$\frac{d}{dt} (E(q(t),p(t))) = 0\,,\quad \frac{d}{dt} (T(q(t),p(t),t)) = 0,$$ then
we have $$E(q(t),p(t))=E_0\,,\quad T(q(t),p(t),t) = T_0,$$ constants
depending on the initial conditions $q(0), p(0)$. This information
is enough to find the solution $q(t), p(t)$ of the system:
\be \label{eq:sol_HO} \begin{cases}
 q(t) = \sqrt{2 E_0} \sin(t-T_0)\\
 p(t) = \sqrt{2 E_0} \cos(t-T_0)
 \end{cases}
\ee
 which can be checked by direct substitution in (\ref{eq:DHO}).

In Fig.\ref{f:HO_parts}, upper panel, we show level surface of conservation law $E = 30$ (on the left),
level surface of dynamical invariant
$T=0$ (middle) and solution trajectory $q(t),p(t)$ (on the right). This solution is actually the intersection
of the level surfaces $E=30$ and
$T=0$; for completeness of presentation, we show together the level surfaces and the solution in Fig.\ref{f:HO_all}
(left panel).

Notice that coordinates $q,p$ are not suitable for a global parametrization of dynamical invariant $T$, (\ref{eq:T_HO}),
 because $\arctan(x)$
is multi-valued. This problem can easily be overcome by using new coordinates $(R,\theta)$:
 \be \label{eq:cov_HO}
 q = R \sin \theta\,,\quad
 p = R \cos \theta
\ee which allow us to rewrite $T$ as 
\be \label{eq:Tnew_HO} T(q,p,t) = \tilde{T}(R,\theta,t) = t - \theta, \ee
 thereby eliminating the ambiguity.
The plot of level surface
$\tilde{T}=0$ in the middle upper panel of Fig.\ref{f:HO_parts} was done using the parameters $(R,t)$.

2. Case $0<\a<2$. Let  $\a/2 = \sin \varphi$, then a dynamical
invariant for the system is known:
 \be \label{eq:D_DHO} D(q,p,t) =
\left(p^2 + q^2 + 2 \,p\, q \sin \varphi \right) \exp(2 \, t \sin
\varphi).\ee
 We need to find a CL in order to determine the solution. Here we simply state the following CL:
\bea \label{eq:C_DHO} C(q,p) &=& \frac{\cos \varphi}{2} \left(p^2 + q^2 + 2\,p\, q \sin \varphi \right) \\
&\times& \exp{\Big[ 2 \tan \varphi \arctan(\frac{q}{p} \sec \varphi
+ \tan \varphi)\Big]}. \nonumber \eea
 The method of construction of this CL is not important right
now, it will be detailed in the next section. In
Fig.\ref{f:HO_parts}, lower panel, we show level surface of
conservation law $C = 30$ (on the left), level surface of dynamical
invariant $D=20$ (middle) and solution trajectory $q(t),p(t)$ (on
the right). This solution corresponds to the intersection of the
level surfaces $C=30$ and $D=20$; for completeness, the level
surfaces and the solution are shown together in Fig.\ref{f:HO_all}
(right panel).

Similar to the case of $\a=0$ dynamical invariant, the CL $C(q,p)$
is not globally defined.  The following change of variables is
needed:
\be \label{eq:cov_DHO} q = R\,\sec \varphi \,\sin (\theta - \varphi ),
  \quad p = R\,\cos \theta \,\sec \varphi, \ee
in order to parameterize globally the CL. The result is \be
\label{eq:Cnew_DHO} C(q,p) = \tilde{C}(R,\theta) = \frac{\cos
\varphi}{2} R^2 \exp(2 \theta \tan \varphi).\ee
 It is important to realize that the two changes of variables (\ref{eq:cov_HO}) and (\ref{eq:cov_DHO}) are suggested by the
form of the respective invariants. Moreover, in the new variables $(R,\theta),$ both $T$ and $C$ take a simpler form (see
(\ref{eq:Tnew_HO}),(\ref{eq:Cnew_DHO})). The reason for it is clear in the case $\a=0$ because $(R,\theta)$ are the
well-known action-angle variables. In the general case, the variables $(R,\theta)$ determine a covering of the original
variables.

The solution of the dynamical system is finally
\bea \label{eq:sol_DHO}
 q(t) &=& \frac{{\sqrt{D}}\,
      \sin (   t\,\cos \varphi  +
       \frac{1}{2} \log (\frac{C}{D}) \cot \varphi -\varphi)}{\cos \varphi \,\exp(
      {t\,\sin \varphi })} \\
 p(t) &=& \frac{{\sqrt{D}}\,\cos (t\,\cos \varphi +
     \frac{1}{2} \log (\frac{C}{D}) \cot \varphi)}{\cos \varphi \, \exp(
    {t\,\sin \varphi })}.
\eea

\subsubsection{Description of a laboratory experiment}

\emph{To check the constancy of the conservation law $C(q,p)$ in a school laboratory experiment,} we give here detailed
description of the possible experiment in the case of sub-critical damping $0 < \a < 2$. Moreover, we provide evidence of
the practical and physical interest of the conservation law $C(q,p)$: it can be used to measure the damping coefficient
$\a$. A simple pendulum oscillating at small amplitudes is easy to construct, consisting of a massive bob attached to the
ceiling by a light (i.e. massless) string or rod. Damping can be easily introduced by attaching to the pendulum rod a sheet
made of a light material. For small amplitudes, the angular frequency of the oscillations is given by $\omega =
\sqrt{g/L}$, where $g \approx 9.8 [m/s^2]$ is the acceleration of gravity and $L$ is the pendulum length. We assume
$\omega$ to be known, though of course it could be measured experimentally. From now on, we choose units where $\omega =
1$, so the damped pendulum equations for small amplitudes reduces to the system (\ref{eq:DHO}), where $q$ is the position
of the oscillating bob and $p$ is its velocity.

With the aid of cheap electronic detection equipment it is possible to measure accurately the time $t$, position $q$ and
velocity $p$ of the oscillating bob at certain instances, namely when the bob passes near a detector. This data is sent to
a computer database and analyzed with a software that comes along with the detection
equipment. As a result one obtains a set of data points of the form%
\be \label{data}(q_n, p_n, t_n)\,,\quad n = 0, \ldots, N \ee%
 With
the aid of only one detector, we can in principle measure data at
instances when $q_n=0$. To wit, we first calibrate the detector by
setting the $q=0$ coordinate as the equilibrium position of the
mass. An experimental realization consists in producing a small
amplitude oscillation and acquiring data points of the form
(\ref{data}). By looking at (\ref{eq:cov_DHO}), these instances
correspond to%
$$\theta_n =
\varphi + n \pi, \quad p_n = (-1)^n R_n,$$%
 where $\varphi$ is the unknown parameter related to the damping coefficient by $\alpha = 2 \sin
\varphi$. From (\ref{eq:Cnew_DHO}) we obtain data values
$$C_n = \frac{\cos \varphi}{2} p_n^2 \exp(2 (\varphi + n
\pi) \tan \varphi ).$$ If $C$ is a conservation law, then $C_n$ must be constant,
independent of $n$. Therefore we should have%
 $$p_n^2 \exp (2 n \pi
\tan \varphi) = \const = p_0^2.$$%
 In practice one can measure $\tan
\varphi$ by finding a linear fit of $\log |p_n|$
versus $n$:%
 \be \label{eq:exp_method_1} \log |p_n| = \log |p_0| - n
\pi \tan \varphi\,, \quad n = 1, \ldots, N. \ee %
Once $\tan \varphi$
has been obtained, the damping coefficient $\a$ is readily computed.

Notice that one could use alternatively the dynamical invariant
$D(q,p,t)$ to compute $\a$, from the data (\ref{data}). Setting $t_0
= 0,$ from (\ref{eq:D_DHO}) we obtain%
 $$D_n = p_n^2 \exp(2 t_n
\sin \varphi) = \const = p_0^2.$$%
 We can measure $\sin \varphi$ by
finding a linear fit of $\log |p_n|$ versus $t_n$:
 \be \label{eq:exp_method_2} \log |p_n| = \log |p_0| - t_n \sin \varphi\,, \quad n
= 1, \ldots, N. \ee Finally, notice that these two ways to measure
the damping coefficient imply a third way: by comparing
(\ref{eq:exp_method_1}) and (\ref{eq:exp_method_2}) we solve for
the data times%
 $$t_n = n \pi/\cos \varphi\,, \quad n = 1, \ldots, N.$$%
In this case, a simple linear fit of $t_n$ as a function of $n$ will
allow us to obtain $\varphi$ and thereby the damping coefficient
$\a.$

\subsubsection{Construction of conservation law}

\emph{To illustrate the procedure of construction of a conservation law}, we take as an example the sub-critically damped
harmonic oscillator, i.e. (\ref{eq:DHO}) with $0 < \a < 2$. Here $N=2$ and $\Delta^1 = p,\quad \Delta^2 = -q - \a p $. The
dynamical system is just $2$-dimensional and we will write it as a vector $(\Delta^1, \Delta^2)^T$. The Theorem requires
the existence of a standard Liouville volume density
$\rho(x^1,x^2)$ satisfying%
\be \label{eq:rho}(\rho \Delta^1)_{,1} + (\rho \Delta^2)_{,2} = 0,\ee%
and does not require the knowledge of conservation laws. In general, a Liouville density, solution of (\ref{eq:rho}), is
interpreted as follows. A small region $\mathcal{R}(t)$ with a volume $V(t)$ in phase space $(x_1,x_2)$, will evolve in
time due to the dynamical system (\ref{eq:DHO}). Then, $\rho$ is defined in such a way that the product $\rho\, V(t)$ is
conserved in time as $\mathcal{R}(t)$ evolves. For the harmonic oscillator, it is well known that the volume of
$\mathcal{R}(t)$ is preserved, i.e., a constant function is a Liouville density. For the damped harmonic oscillator, a
direct check shows that a Liouville density is $\rho(q,p) = \left(q^2+p^2+2\a \,q\,p\right)^{-1}.$ With this information we
just need to solve (\ref{eq:Ham_Matrix}) for $H$:
\be \label{eq:Ham_Matrix}\left(
      \begin{array}{cc}
        0 &1 \\
        -1 & 0 \\
      \end{array}
    \right) \left(
              \begin{array}{c}
                H_{,1} \\
                H_{,2} \\
              \end{array}
            \right) = \left(
                        \begin{array}{c}
                           \rho\, \Delta^1 \\
                          \rho\, \Delta^2 \\
                        \end{array}
                      \right).
\ee
The answer
can be obtained by direct integration (see \cite{BK09_1} for more details):%
\bea H(q,p) = -\left( \frac{q\,\alpha \,\arctan (\frac{2\,p +
q\,\alpha }
         {{\sqrt{-\left( q^2\,\left( -4 + {\alpha }^2 \right)  \right) }}})}
       {{\sqrt{-\left( q^2\,\left( -4 + {\alpha }^2 \right)  \right) }}}
       \right)+\nn\\
      + \frac{\log (p^2 + q^2 + p\,q\,\alpha )}{2} \eea%
The conservation law $C(q,p)$ given by Eq.(\ref{eq:C_DHO}) is a
function of  $H$, chosen for its nice form:  $$C = \frac{\cos
\varphi}{2} \exp(2 H).$$

\section{Triad}
As it was shown above, the notion of  dynamical invariant is an important tool for constructing new physically relevant
conservation laws that can afterwards be studied in a simple laboratory experiment. In this section we would like to use
this approach to prove integrability of a complex triad with dynamical system (\ref{dyn3waves}). Of course, integrability
of (\ref{dyn3waves}) is a well-known fact (e.g. \cite{book-triad}). However, the explicit solution of (\ref{dyn3waves})
is usually written for a particular case, namely, when the dynamical phase -- a phase combination corresponding to the chosen resonance conditions --
is either zero or constant
(\cite{LongHigGill67}, p.132, Eq.(6.7); \cite{ped}, p.156, Eq.(3.26.19), etc.).

 On the other hand, it is well known that
dynamical phases play a substantial role in the dynamics of resonant clusters, e.g. \cite{tsyt}, and their effect can  easily be
observed  in numerical simulations, \cite{BK09_2}. This was our motivation for constructing first an explicit
solution in the amplitude-phase presentation, for (\ref{Triad-A-Ph}). Thus, (\ref{dyn3waves}) is used for a
preliminary check of our method.

Another important point is the following. As it was shown in the papers \cite{Ly02a,Ly02b,LH04}, an elastic pendulum with
suitably chosen parameters can be used as a mechanical model of a resonant triad, and the results can be applied for the
description of large-scale motions in the Earth's atmosphere. In fact, this simple mechanical model can be used for a
laboratory study of dynamical characteristics of primary clusters in an \emph{arbitrary} system with cubic Hamiltonian.
This is why the explicit analytical formulas for all dynamically relevant parameters, given below, are important.

\subsection{Integrability}

In this case the system can be reduced to $N=4$ (see \cite{BK09_1} for more details), the Theorem on $(N-2)$--integrability
can be applied and we obtain the following CL:
 \be \nonumber \label{triad-CLs_new} H_T = \operatorname{Im}(B_1 B_{2} B_{3}^*)\,,\ee
which is the canonical Hamiltonian for this case and can, of course, be written out directly.
A dynamical invariant for
this system was originally presented in \cite{BK09_1}, in terms of the three real roots $R_1<R_2<R_3$ of the cubic
polynomial
$$x^3 + x^2 = 2/27
-(27 H_T^2 - $$
$$ (I_{13}+I_{23})(I_{13}-2
I_{23})(I_{23}-2 I_{13}))/27 (I_{13}^2 - I_{13}I_{23} + I_{23}^2)^{3/2}\,,$$ 
but these roots' dependence on the coordinates or the CLs was not made explicit. Moreover,
the explicit solution for the amplitudes $C_j$ and
phases $\theta_j$ in the amplitude-phase representation $B_j = C_j \exp\left(i \theta_j\right)$ was
not provided. Here we improve the form of
dynamical invariant and also produce explicit and useful expressions for the full solution,
based on the trigonometric representation of the
three real roots in the so-called \emph{Casus Irreducibilis}.

\subsection{Amplitude-phase representation}

Sys.(\ref{dyn3waves}) in the standard amplitude-phase representation $B_j= C_j\exp(i \theta_j)$ reads:
 \bea \label{Triad-A-Ph}
\begin{cases}
\dot{C}_1=Z C_2C_3 \cos \varphi,\\
\dot{C}_2=Z C_1C_3 \cos \varphi,\\
\dot{C}_3=- Z C_1C_2 \cos \varphi,\\
\dot{\varphi}= -Z\,H_T (C_1^{-2}+C_2^{-2}-C_3^{-2}).
\end{cases}
\eea
where $\varphi=\theta_1+\theta_2-\theta_3$ is the dynamical phase.  The conservation laws (\ref{laws3waves}) do not change
their form in the new variables:
$$
 I_{23}=C_2 ^2 + C_3^2,  \  \ I_{13}= C_1^2 + C_3^2,
$$
but
 the Hamiltonian $H_T$  reads now
\be \label{eq: H_T} H_T=C_1 C_2 C_3 \sin \varphi.
 \ee
Let us introduce new variables:%
\be \label{rho} \rho = I_{23}/I_{13}\ee
 and $\alpha \in [0,\pi]$
defined by
\be \label{cos_alp}\cos \alpha = \frac{\left( -2 + 3\,\rho  +
3\,{\rho }^2 - 2\,{\rho }^3 \right) \,
     {{{I}_{13}}}^3 - 27\,{{{H}_T}}^2}{2\,
    {\left( 1 - \rho  + {\rho }^2 \right) }^{\frac{3}{2}}\,
    {{{I }_{13}}}^3}\,.\ee%
Notice that $|\cos \alpha| \leq 1$ for dynamically accessible system's configurations. Indeed, the use of intermediate
variables $p = {2\, {\left( 1 - \rho + {\rho }^2 \right) }^{\frac{3}{2}}}$ and $q = -2 + 3\,\rho  + 3\,{\rho }^2 - 2\,{\rho
}^3\,,$ allows one to conclude immediately that $p \geq 0$ and $p \geq \left| q \right| \quad \forall \, \rho\,.$ Both
inequalities become equalities if $\rho = 0$ or $\rho = 1.$ This yields
$$
\left(H_T^2\right)_{\mathrm{max}} = \left(\frac{I_{13}}{3}
\right)^{3} \left(p + q\right) \geq 0\,, \quad \mbox{for} \quad \cos
\alpha = -1,
$$
and
$$
(\cos \alpha)_{\mathrm{max}} = q/p \leq 1\,, \quad \mbox{for} \quad
H_{T} = 0,
$$
where $(H_T^2)_{\mathrm{max}}$ and $(\cos \alpha)_{\mathrm{max}}$ are maximum values of $H_T^2$ and $\cos \alpha$ correspondingly.

Now, the solution of (\ref{Triad-A-Ph}) is obtained in terms of
Jacobian functions with modulus
\be \label{eq: mu}\mu = \cos \left(\frac{\alpha }{3} + \frac{\pi
}{6}\right)/
      \cos \left(\frac{\alpha }{3} - \frac{\pi }{6}\right)\,,\ee%
and period
 \be \label{eq: period} T = \frac{{\sqrt{2}}\,3^{\frac{1}{4}}\,
    {K}\left(\mu\right)}{Z\,
    {\left( 1 - \rho  + {\rho }^2 \right) }^{\frac{1}{4}}\,
    {\sqrt{\cos (\frac{\alpha }{3} - \frac{\pi }{6})}}\,
    {\sqrt{{{{I}}_{13}}}}}\,, \ee
where $K(\mu)$ is the complete elliptic integral of the first kind.

\subsection{Solutions for amplitudes}

We present explicit expressions for the amplitude squares. The convention used here is that the amplitudes are positive,
which is the generic situation when $H_T \neq 0\,.$ In this convention, when $H_T = 0$ the individual phases have
discontinuities in time to account for the amplitudes' sign
changes. The amplitude squares are proportional to the modes' energies and can be of great use for physical applications:%
\bea \label{sol3wavesGeneral}
\begin{cases}
C_1^2(t) = -\mu\,\left(\frac{2 K(\mu)}{Z\,T}\right)^2 {{\mathbf{sn}^2}\left(2\,K(\mu )\,
             \frac{{(t-t_0)}}{{T}},\mu \right)}\\ \quad \quad \quad + \frac{I_{13}}{3}\left({2 - \rho} + 2\,{\sqrt{1 - \rho  + {\rho }^2}}\,
      \cos \left(\frac{\alpha}{3}\right)\right)\,,\\
C_2^2(t) = -\mu\,\left(\frac{2 K(\mu)}{Z\,T}\right)^2 {{\mathbf{sn}^2}\left(2\,K(\mu )\,
             \frac{{(t-t_0)}}{{T}},\mu \right)}\\ \quad \quad \quad + \frac{I_{13}}{3}\left({2 \rho - 1} + 2\,{\sqrt{1 - \rho  + {\rho }^2}}\,
      \cos \left(\frac{\alpha}{3}\right)\right)\,,\\
C_3^2(t) = \mu\,\left(\frac{2 K(\mu)}{Z\,T}\right)^2 {{\mathbf{sn}^2}\left(2\,K(\mu )\,
             \frac{{(t-t_0)}}{{T}},\mu \right)} \\ \quad \quad \quad + \frac{I_{13}}{3}\left({\rho + 1} - 2\,{\sqrt{1 - \rho  + {\rho }^2}}\,
      \cos \left(\frac{\alpha}{3}\right)\right)\,,
\end{cases}
\eea 
where $\mathbf{sn}(\cdot,\mu)$ is  Jacobian elliptic function and $t_0$ is given in terms of the initial conditions for the amplitudes $C_1^2(0),
C_2^2(0), C_3^2(0)$ and $t_0$ is defined by the initial conditions as:

\begin{equation} \label{t0}
t_0 = \textrm{sign}(\cos \varphi(0)) \,\frac{T}{2\,K(\mu)}\times
 F\left(\arcsin \sqrt{x_0},\mu \right)\,,
 \end{equation}
where
\be \label{x0}
x_0=\frac{\cos \left(\frac{\alpha }{3}\right)}{\sqrt{3}\, \cos \left(\frac{\alpha }{3} + \frac{\pi }{6}\right)} +
\frac{Z^2\,T^2\,(C_3^2(0) - C_2^2(0) - C_1^2(0))}{12\,\mu\,K(\mu)^2}
\ee
and $F(\cdot,\mu)$ is the elliptic integral of the first kind.

 Notice that each equation in (\ref{sol3wavesGeneral}) is a sum of two terms where the left terms are time-dependent and the right terms are not.
Each right term, for instance
$$
\frac{I_{13}}{3}\left({2 - \rho} + 2\,{\sqrt{1 - \rho  + {\rho }^2}}\, \cos \left(\frac{\alpha}{3}\right)\right)
$$
 can be written explicitly as a function of conserved quantities $I_{13}, \ I_{23}, \ H_T$
   (expressions for $\rho$ and $\alpha$ are given by
   (\ref{rho}),(\ref{cos_alp}))  and is, therefore, defined by
   the initial conditions.

   The same is true for $\mu$ and
   $T$ as it follows from (\ref{eq: mu}) and (\ref{eq: period}). In particular, one can use the equations in  (\ref{sol3wavesGeneral}) to determine
   the minimum and maximum accessible values of each amplitude (using the fact that $\mathbf{sn}^2$ oscillates between $0$ and $1$). The
   characteristic energy variation of any resonant mode $E_{mode},$ between these minimum and maximum values, has a very
   simple form: $E_{mode}(t) \sim \mathbf{sn}^2(k\,t,\mu).$

\subsection{Solution for dynamical phase}

The dynamical phase  satisfies an evolution equation:%
 \be \label{DynPhaseT} \dot{\varphi}= - Z\,H_T
(C_1^{-2}+C_2^{-2}-C_3^{-2}).
 \ee
The solution for the dynamical phase \textbf{cannot} be obtained by
simply replacing the solution for the amplitudes in the Hamiltonian
$H_T=C_1 C_2 C_3 \sin \varphi$ and solving for $\varphi$. The reason
is that non-zero $\varphi$ generically evolves between $0$ and
$\pi$, crossing the value $\varphi = \pi/2$ periodically. This
implies that $\sin^{-1}$ is double-valued and thus it is not
possible to obtain $\varphi$ in a unique way.

Another way to obtain the solution for dynamical phase might be
integrating (\ref{DynPhaseT}) in time, using the solution
 for the amplitude squares,
(\ref{sol3wavesGeneral}), but this way is also rather involved. On the other hand, some simple considerations allow us to find an analytical
expression for the dynamical phase. Indeed, let us rewrite (\ref{Triad-A-Ph}), taking into account that $\frac{d}{dt}C_1^2 = 2 C_1 \dot{C}_1$ and
$H_T=C_1 C_2 C_3 \sin \varphi$:
$$\frac{d}{dt}C_1^2 = 2 Z C_1 C_2 C_3 \cos \varphi = 2 Z H_T \cot \varphi\,.$$
This equation can be solved for $\varphi$ in each of the disjoint domains $(0,\pi)$ and $(-\pi,0)$:
$$\varphi(t) = \textrm{sign}(\varphi(0))\,\mathrm{\mathbf{arccot}}\left(\frac{\textrm{sign}(\varphi(0))\,\frac{d}{dt}C_1^2}{2\, Z\, H_T}\right)\,,$$
using the convention that the function $\mathrm{\mathbf{arccot}}$ takes values on $(0,\pi).$ Using solution (\ref{sol3wavesGeneral}) together
with the identity $\mathrm{\mathbf{sn}}'(x,\mu) = \mathrm{\mathbf{cn}}(x,\mu) \mathrm{\mathbf{dn}}(x,\mu)$ we arrive at an explicit expression
for the dynamical phase:
 \be \label{eq: phase} \varphi(t) = \textrm{sign}(\varphi_0)\,\mathrm{\mathbf{arccot}}\left(-\frac{\mu}{|H_T|} \left(\frac{2 K(\mu)}{Z\,T}\right)^3
y \right)\,,
 \ee
where
\be
 y=\mathrm{\mathbf{sn\,cn\,dn}} \left(2\,K(\mu )\,
             \frac{{(t-t_0)}}{{T}},\mu \right)               \nonumber
\ee
and $\mathrm{\mathbf{sn\,cn\,dn}}(\cdot,\mu) \equiv \mathrm{\mathbf{sn}}(\cdot,\mu) \,\mathrm{\mathbf{cn}}(\cdot,\mu) \,\mathrm{\mathbf{dn}}(\cdot,\mu)\,. $

 The restriction to the domain $\varphi \in (-\pi,0)\cup(0,\pi)$
 is quite general: if $\varphi$ is initially in the domain $(n\,\pi, (n+1)\,\pi)\,, n \in \mathbb{Z},$ one can take
 $\varphi$ to either $(-\pi,0)$ or $(0,\pi)$ by an appropriate shift of $2\, m\,
\pi\,, m \in \mathbb{Z}\,,$ without changing the evolution equations. Due to its special dynamics, the phase will remain in the domain where it
was initially.

We can simplify (\ref{eq: phase}) using nome $q$ of
 elliptic function defined as
$$
q= \exp \left(- \pi \frac{K'(\mu)}{K(\mu)}\right).
$$
Since
\bea
\frac{2K(\mu)}{\pi} \mathbf{sn}\left(2\,K(\mu )\,
             \frac{{(t-t_0)}}{{T}}, \mu\right) = \nonumber \\
             \frac{4}{\mu}\sum_{n=0}^{\infty}\frac{q^{(2n+1)/2}}{1-q^{2n+1}}\sin(2n+1)\pi \frac{{(t-t_0)}}{{T}},
\eea

one can compute $\mathrm{\mathbf{sn\,cn\,dn}}(x,\mu)$ using the fact that $\mathbf{sn}' = \mathbf{cn\,dn}$. Explicit expression would be then a product of two infinite sums \cite{K09b} and can be rewritten as a Fourier series. Since the nome $q$ is an explicit function of the initial conditions, to
have Fourier representations of our dynamical variables might be useful regarding approximate solutions.

\subsection{Dynamical invariant}
Below we present a dynamical invariant for (\ref{Triad-A-Ph}) which has been used for the constructing
 the solution (\ref{sol3wavesGeneral}). Recall that a dynamical invariant depends on time, amplitudes
and phases: $S(t,C_1,C_2,C_3,\varphi)$, with the property that it is a constant along any solution of the dynamical system:
$S(t,C_1(t),C_2(t),C_3(t),\varphi(t)) = \mathrm{const.} \, \forall \, t\,.$ Generically, only local expressions can be
obtained for a dynamical invariant, due to the multi-valuedness of the inverse functions involved. In this particular case,
however, since we know the period of any trajectory, this multi-valuedness can be eliminated partially by patching
appropriately local expressions yielding
\bea  \label{triad-dyn}
S(t,C_1,C_2,C_3,\varphi)
  =  t - \left\lfloor\frac{2(t-t_0)+T}{2\,T}\right\rfloor \, T \nonumber\\ +
\, \frac{(-1)^{\left\lfloor\frac{2(t-t_0)+T}{T}\right\rfloor}\,T}{2\,K(\mu)}\,
 F\left(\arcsin \sqrt{x_t},\mu \right)\,,
\eea
where $\lfloor \cdot\rfloor$ is the floor function and
$x_t$ can be obtained from the expression (\ref{x0}) for $x_0$ by substituting $C_j(t)$ instead of $C_j(0)$, for all $j=1,2,3.$

This dynamical invariant satisfies
$S(t,C_1(t),C_2(t),C_3(t),\varphi(t)) = t_0 \quad \forall t\,,$ where $t_0$ is given in equation (\ref{t0}), and is an
improvement of the corresponding formula presented in \cite{BK09_1}.

\begin{figure*}
\begin{center}
\includegraphics[width=5cm,height=5cm,angle=270]{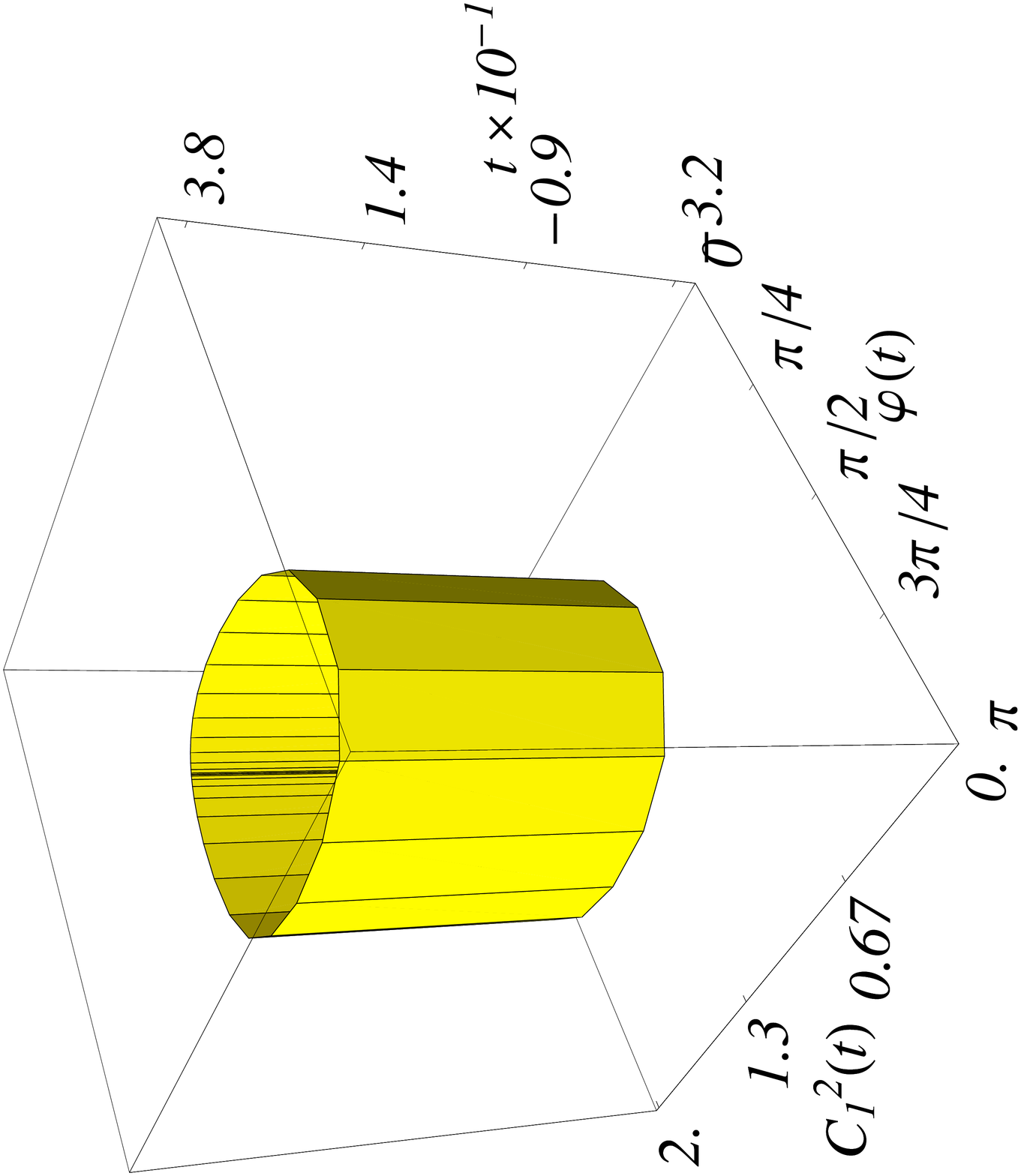}
\includegraphics[width=6cm,height=6cm,angle=270]{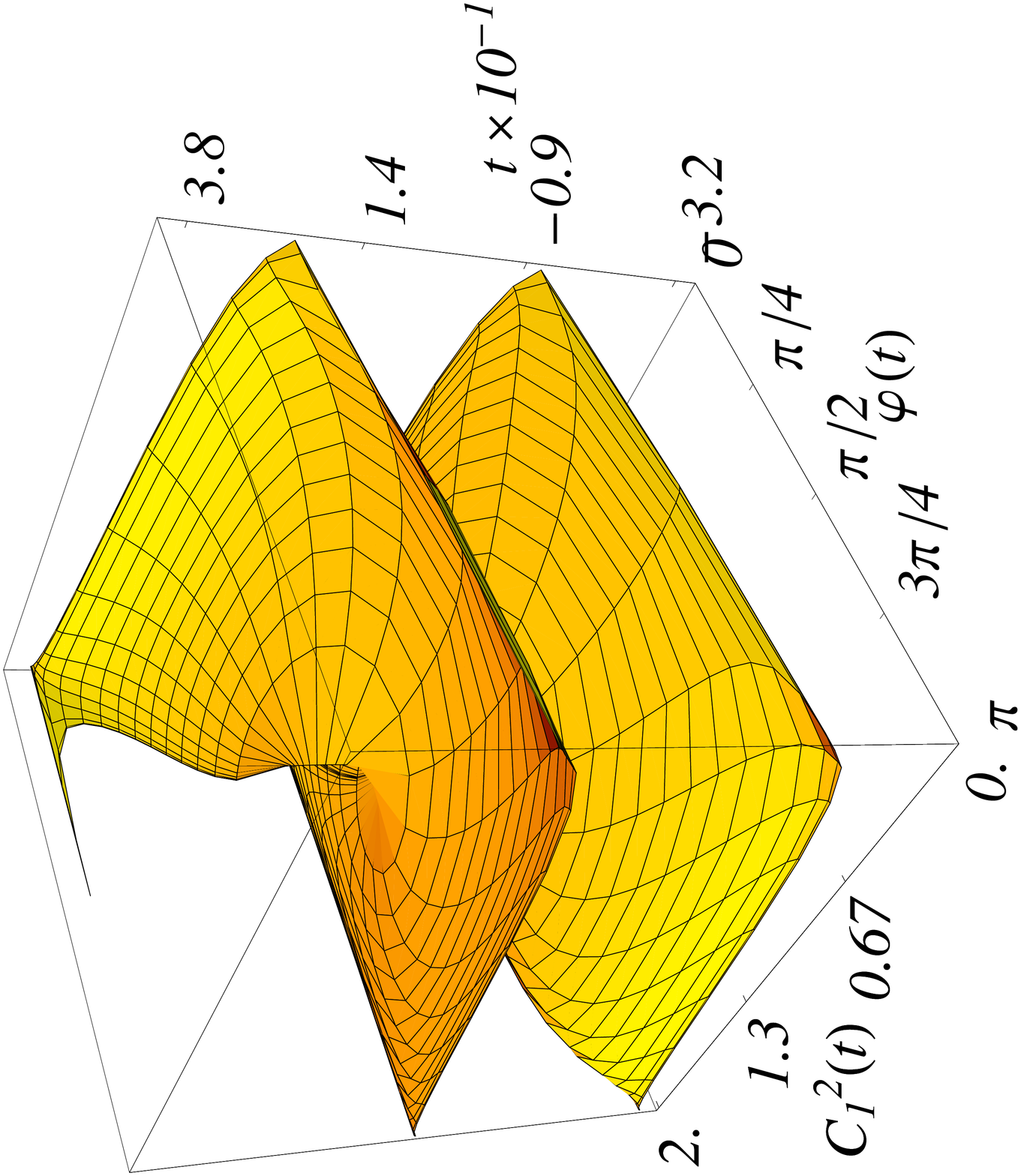}
\includegraphics[width=5cm,height=5cm,angle=270]{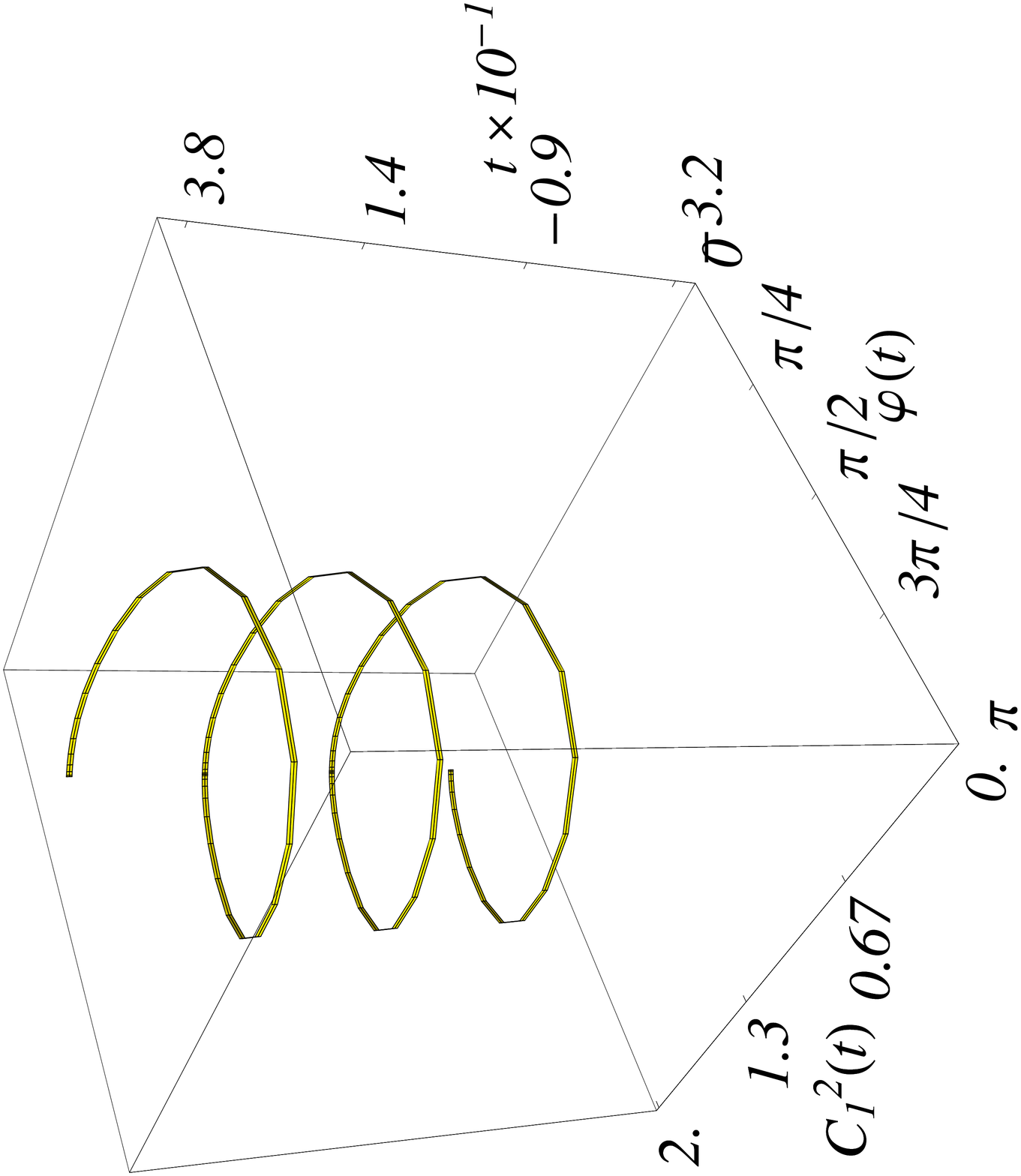}
\includegraphics[width=5cm,height=5cm,angle=270]{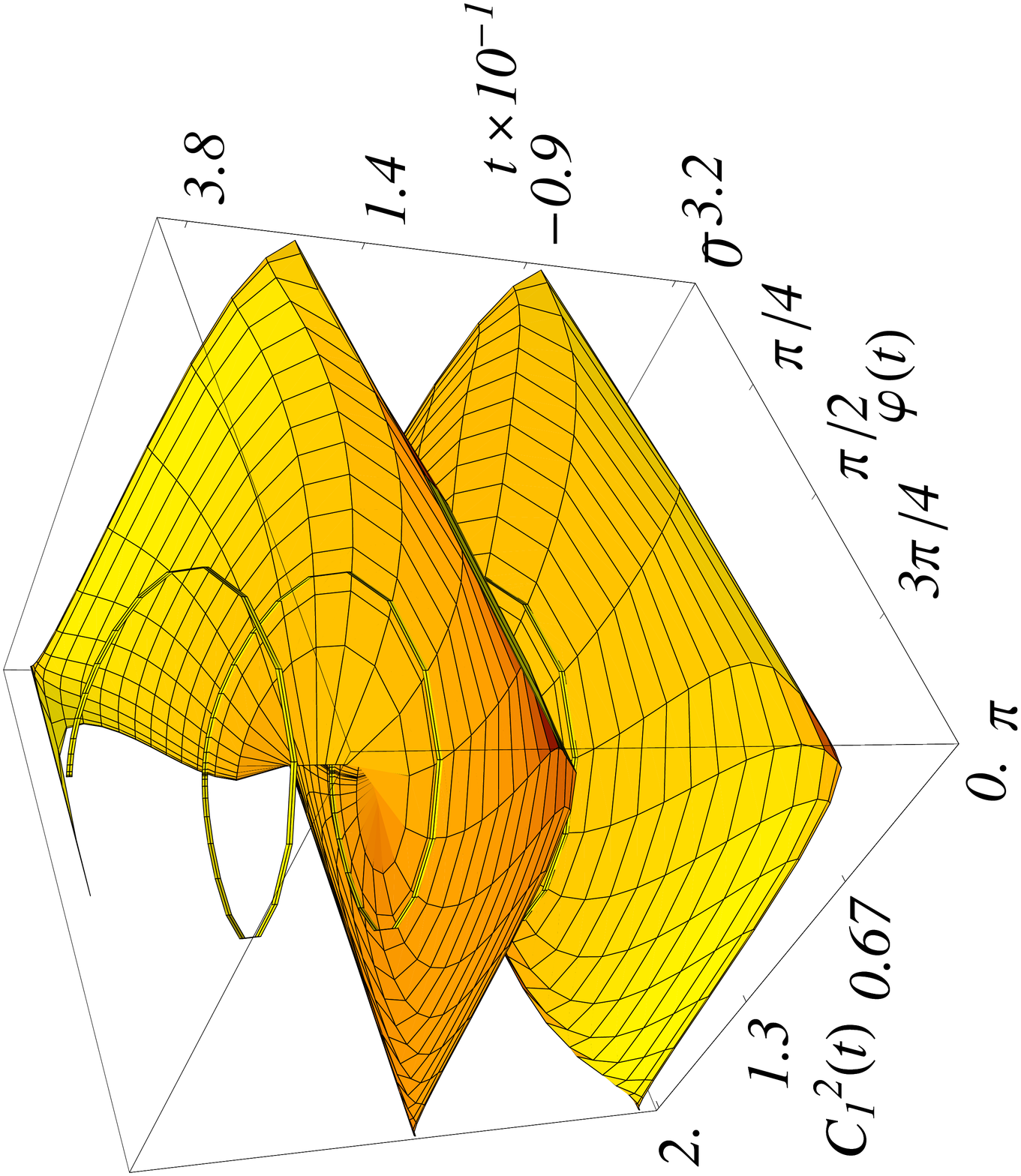}
\includegraphics[width=6cm,height=6cm,angle=270]{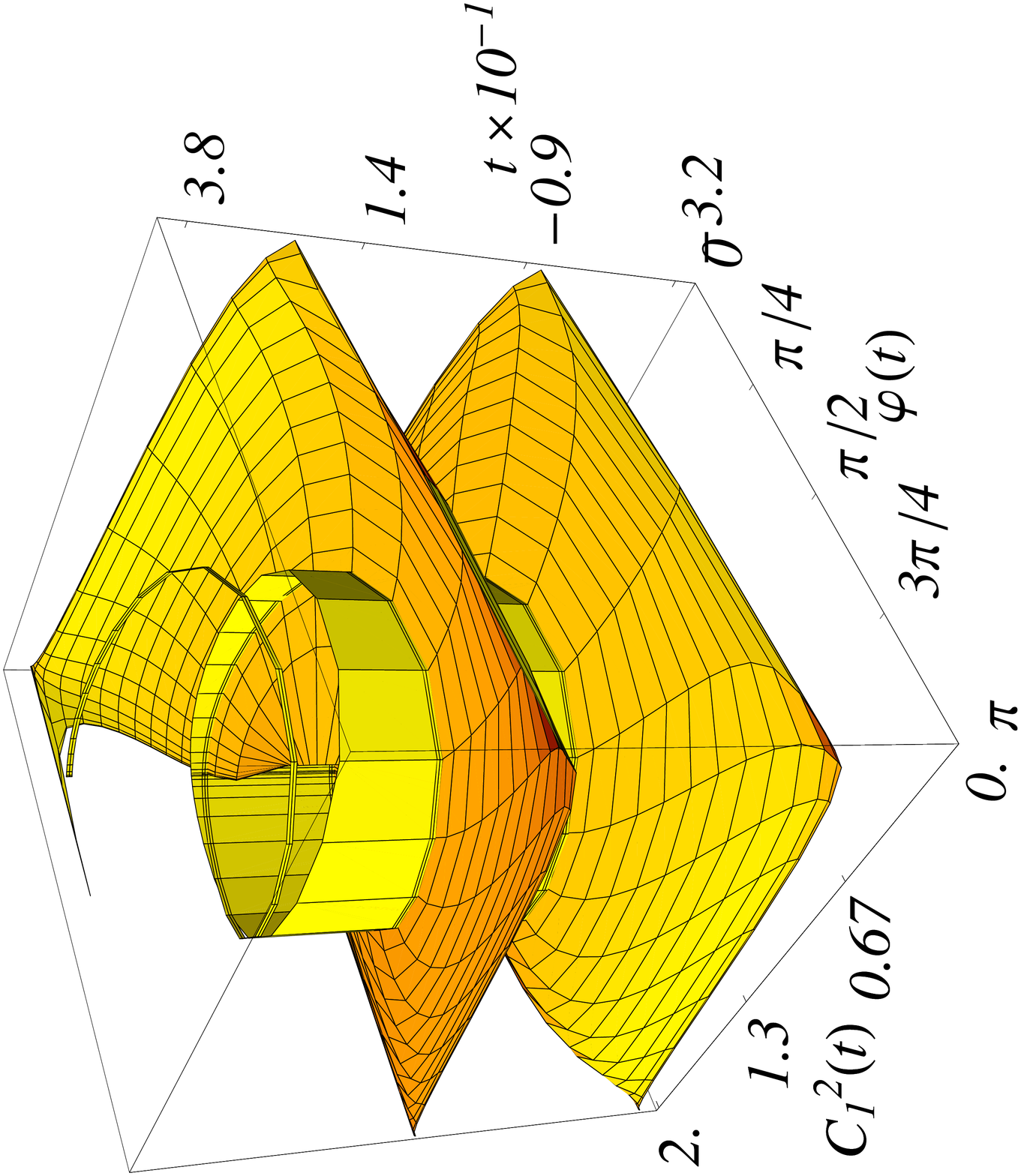}
\includegraphics[width=5cm,height=5cm,angle=270]{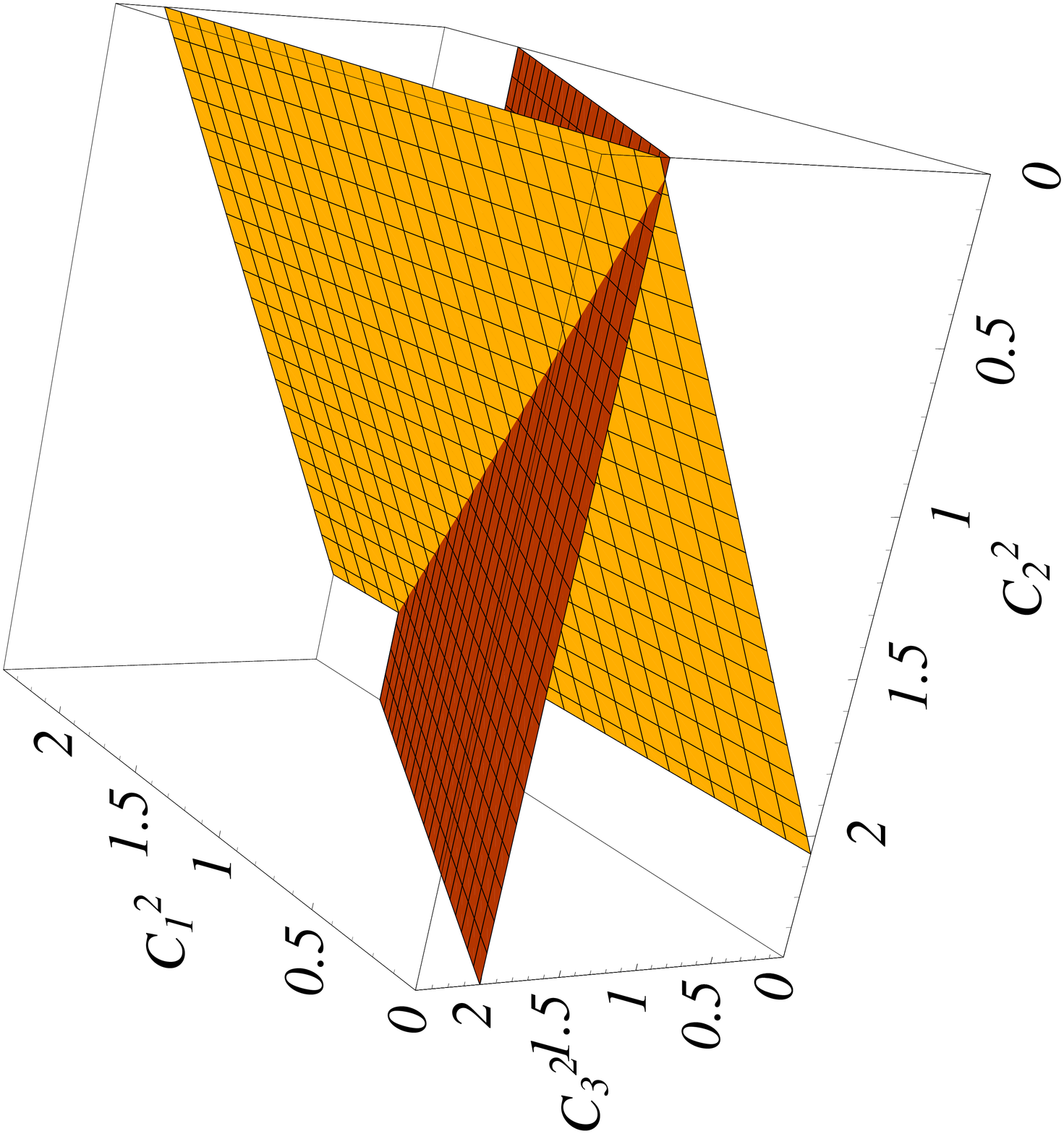}
\end{center}
\caption{\label{f:Triad} Color online. \textbf{Upper panel:} Triad system in coordinates $(C_1^2, \varphi, t)$, with fixed values of the
Manley-Rowe conservation laws: $I_{13} = 2.00$ and $I_{23}=2.06\,.$ \textbf{Left:} level surface of conservation law $H_T = 0.763$, Eq.(\ref{eq:
H_T}). \textbf{Middle:} level surface of dynamical invariant $S = 2.69$, Eq.(\ref{triad-dyn}). \textbf{Right:} solution trajectory
$(C_1^2(t),\varphi(t))$, Eqs.(\ref{sol3wavesGeneral}),(\ref{eq: phase}), corresponding to $H_T=0.763, S=2.69$. \textbf{Lower panel:}
\textbf{Left:} combined plot of level surface of dynamical invariant $S = 2.69$ and solution trajectory $(C_1(t)^2,\varphi(t))$. \textbf{Middle:}
combined plot of level surface of conservation law $H_T=0.763$, level surface of dynamical invariant $S = 2.69$ and solution trajectory
$(C_1^2(t),\varphi(t))$. Notice the general property that the intersection of the level surfaces of $H_T$ and $S$ \emph{is} the solution
trajectory. \textbf{Right:} combined plot of Manley-Rowe conservation laws $I_{13} = 2.00$ and $I_{23}=2.06\,,$ in coordinates
$(C_1^2,C_2^2,C_3^2)\,.$ }
\end{figure*}

In Fig. \ref{f:Triad}, upper panel, we show, for fixed $I_{13} = 2.00$ and $I_{23} = 2.06$: isosurface of conservation law
$H_T = 0.763$, isosurface of dynamical invariant $S = 2.69$, solution trajectory and combined plot, in the domain $(C_1,
\varphi, t).$

In the middle of the upper panel, the surface $S = 2.69$ is a helicoidal surface revolving around a vertical axis. This
axis is the surface's natural interior boundary: the constant-in-time trajectory corresponding to the highest possible
value of $|H_T|$ for given $I_{13}, I_{23}$ (obtained from the condition $\cos \alpha = -1$). In the present case, the
highest possible value of $|H_T|$ is $1.114.$ This trajectory is physically interpreted as `maximum interference', due to
the fact that the modes do not interact. The dynamical phase is constant: $\varphi(t) = \pi/2 \, \forall \,t,$ and all
amplitudes are constant as well: from the condition $\mu=0$ and equations (\ref{sol3wavesGeneral}), we obtain in this case:
$C_1^2(t) = 1.33, \, \forall \,t\,.$  The exterior boundary of the surface is the piecewise continuous trajectory
corresponding to the limit $H_T = 0\,:$ in this limit the surface becomes non-differentiable at the `corners' $C_1^2 = 0,
I_{13},\,\varphi=0,\pi,$ due to the fact that the dynamical phase $\varphi$ is only piecewise continuous for $H_T = 0$.
This trajectory corresponds to the usual case treated in textbooks, when amplitudes are considered real and individual
phases vanish.

\begin{figure}
\begin{center}
\includegraphics[width=5cm,height=6cm,angle=270]{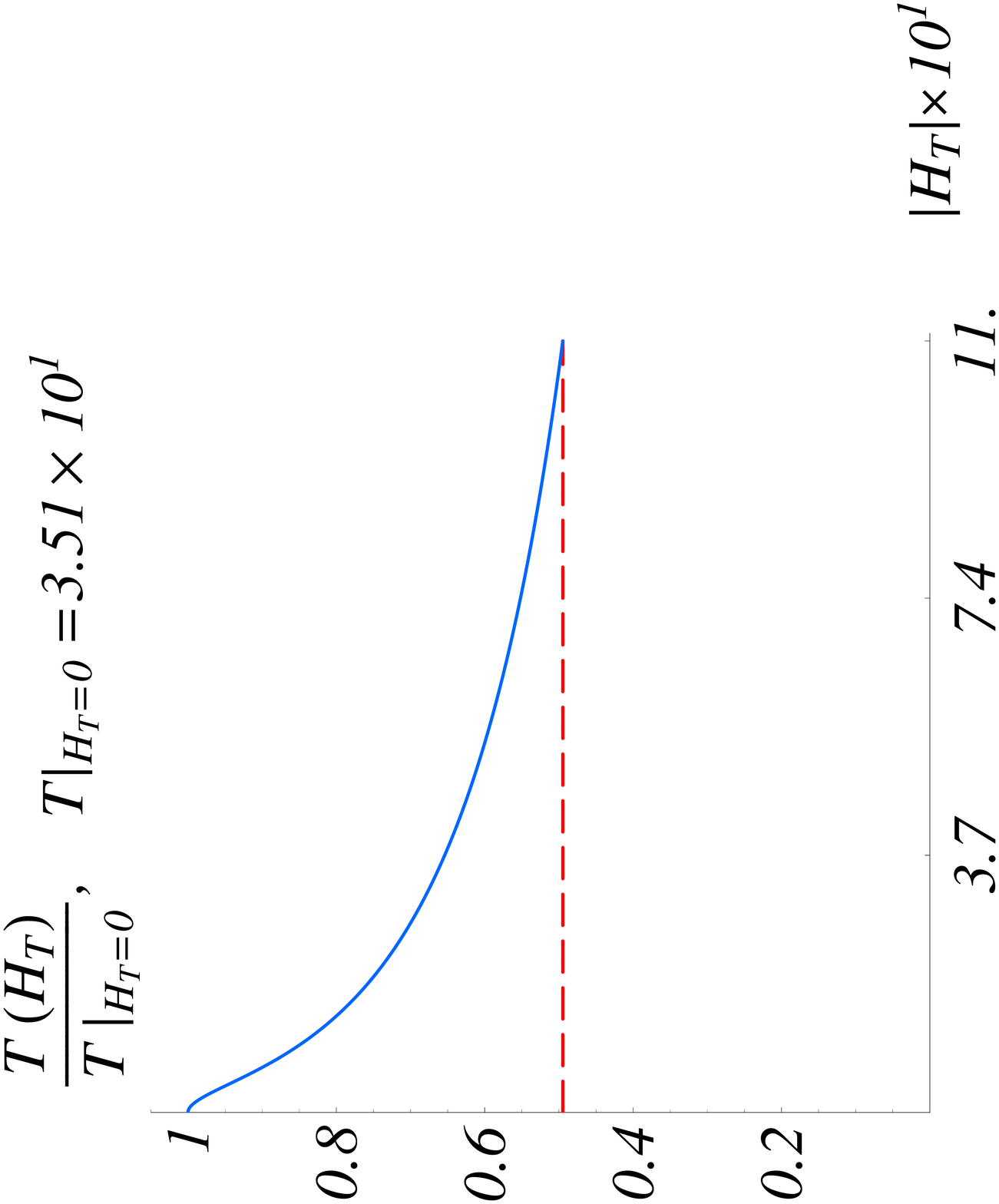}
\end{center}
\caption{\label{f:T vs. H_T} Color online. With fixed values of the Manley-Rowe conservation laws: $I_{13} = 2.00$ and $I_{23}=2.06\,,$ plot of
period $T$ from Eq.(\ref{eq: period}) as a function of $H_T$, normalized with respect to the period at $H_T=0$. The decreasing character is a
generic feature of this function.}
\end{figure}

By looking at this figure we notice that the period $T$ decreases with increasing $H_T$: the trajectories closer to the
exterior boundary are more elongated than the trajectories closer to the interior boundary. In fact, from formula (\ref{eq:
period}) one can prove this property analytically. In Fig. \ref{f:T vs. H_T} we plot the period $T$ as a function of $H_T$.
We observe in this case a reduction of the period by a factor $0.5$ when $H_T$ is changed from 0 to $H_{\mathrm{max}} =
1.114\,.$

In Fig. \ref{f:Triad}, lower panel, on the left and in the middle, combined plots are shown to clarify that the solution
trajectory is the intersection of the isosurfaces of Hamiltonian and dynamical invariant. On the right, we show a combined
plot of level surfaces of Manley--Rowe conservation laws $I_{13}$, $I_{23}$ in the domain $(C_1^2,C_2^2,C_3^2)$.

\subsection{Special case $H_T = 0$}

Direct substitution shows that if we put $H_T=0$, then new modulus
and period take the form
$$\mu = \rho \quad \mbox{and} \quad T = \frac{2 K(\rho)}{Z \sqrt{I_{13}}}$$
correspondingly, while the solutions for the
amplitude squares read%
 \bea \label{sol3wavesReal_Squares}%
\begin{cases}
\Tilde{C}_1^2(t) = \mathrm{dn}^2(\left(t - {t_0} \right) \,Z\,
    {\sqrt{{{{I} }_{13}}}},
   \rho)\,
  {{{{{{I} }}_{13}}}}\\
\Tilde{C}_2^2(t) = \mathrm{cn}^2(\left(t - {t_0} \right) \,Z\,
    {\sqrt{{{{I} }_{13}}}},
   \rho)\,
  {{{{{{I} }}_{23}}}}\,
  \\
\Tilde{C}_3^2(t) = \mathrm{sn}^2(\left(t - {t_0} \right) \,Z\,
    {\sqrt{{{{I} }_{13}}}},
   \rho)\,
  {{{{{{I} }}_{23}}}}\,.
\end{cases}
\eea%
As for the dynamical phase, from Eq.(\ref{eq: phase}) it is seen that in the limit $H_T \to 0$ it behaves as a step function, jumping from $0$ to
$\pi\,\mathrm{sign}(\varphi(0))$:
$$\widetilde{\varphi}(t) = \frac{\pi\,\mathrm{sign}(\varphi(0))}{2} \left(1 - (-1)^{\left\lfloor\frac{2 (t-t_0)}{T}\right\rfloor}\right)\,.$$


To understand the meaning of this behaviour, notice that the Hamiltonian $H_T$ is vanishing for $\varphi=n\,\pi, \, n \in
\mathbb{Z}.$ The abrupt jumps of the dynamical phase is due to the jumps of the individual phases (solution not shown).
These jumps replace the changes of sign of the modes' amplitudes in the usual textbook descriptions.

As it was shown in \cite{BK09_2}, initial dynamical phase not in $\mathbb{Z}\,\pi$ substantially affects the magnitudes of
resonantly interacting modes during the evolution, not only in a triad but also in a butterfly. This fact might have
important implications (see \cite{BK09_2}, Discussion), for instance, for interpreting results of numerical simulations and
for performing laboratory experiments.

\section{Generic clusters}

It is well-known (see, for instance, \cite{PRL94,K06-1,KK07,KL-06,KL-07,all08}, etc.) that in  three-wave resonance systems
the most frequently met clusters are isolated triads or clusters consisting of two variously connected triads. Below we
classify all possible two-triad clusters - kite, butterfly and ray - and show how to construct new CLs making use of the
notion of dynamical invariant. In the last subsection, another method is  briefly outlined which was presented in
\cite{Ver68a,Ver68b} and allows one to prove, in some cases, integrability of bigger clusters.

\subsection{Kite}
A kite consists  of two  triads  $ a $ and $ b, $ with wave
amplitudes
 $ B_{ja}, \ B_{jb}, $ $j=1,2,3$,  connected {\it via} two common
 modes. One can point out 4 types of
 kites according to  the properties of connecting modes: PP-PP, AP-PP, AP-AP and AA-AA kites. In this section,
PP-PP-kite
 with
 $ B_{1a}=B_{1b} $ and $ B_{2a}=B_{2b}$ is taken as a representative example. Its dynamical system reads:
\bea \label{PP-PP}
\begin{cases}
\dot{B}_{1a}=  B_{2a}^* (Z_a B_{3a} +   Z_b B_{3b}) \,, \\
\dot{B}_{2a}=  B_{1a}^* (Z_a B_{3a} +  Z_b B_{3b}) \,, \\
\dot{B}_{3a}=  - Z_{a} B_{1a} B_{2a} \,,\    \dot{B}_{3b}=  - Z_{b}
B_{1a} B_{2a}\ .
\end{cases}
 \eea%
 \begin{figure}[h]
\begin{center}\includegraphics[width=3cm,height=0.75cm]{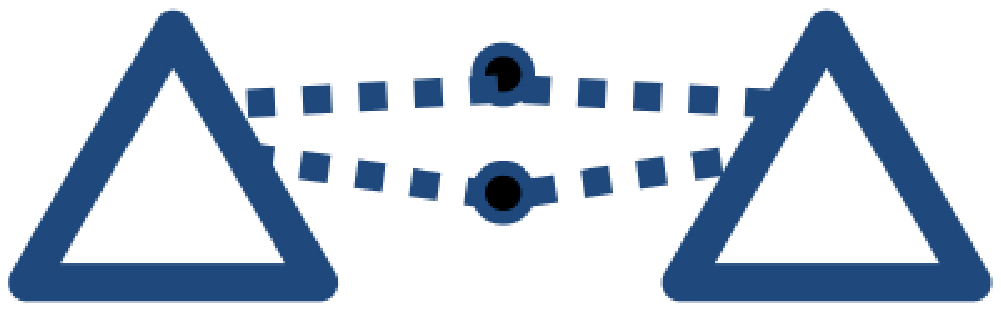}
\end{center}
\caption{\label{NB-diag-PP-kite} NR-diagram for PP-kite.}
\end{figure}
It has 5 conservation laws (2 linear, 2 quadratic, 1
cubic):%
 \bea \nonumber \label{kite-CLs}
 \begin{cases}\nonumber
 L_{R}= \operatorname{Re}(Z_b B_{3a} - Z_a B_{3b}), \ L_{I}= \operatorname{Im}(Z_b B_{3a} - Z_a B_{3b})\,,\\
 I_{1ab}= |B_{1a}|^2 + |B_{3a} |^2 + |B_{3b}|^2\,,\\
 I_{2ab}= |B_{2a}|^2 + |B_{3a} |^2 + |B_{3b}|^2\,,\\
 H_K = \operatorname{Im}(Z_a B_{1a} B_{2a} B_{3a}^*+Z_b B_{1a} B_{2a}
B_{3b}^*)).
 \end{cases}
 \eea %

It has a dynamical invariant that is essentially the same as for a triad, $ S$, after replacing
 $$Z = Z_a+Z_b, \ \   H_T = H_K
(Z_a^2+Z_b^2)/Z^3,$$
 $$I_{13}= I_{1ab} (Z_a^2+Z_b^2)/Z^2 - (L_R^2 +
L_I^2)/Z^2,$$
 $$I_{23}= I_{2ab} (Z_a^2+Z_b^2)/Z^2 - (L_R^2 +
L_I^2)/Z^2.$$

\subsection{Butterfly}
A  \emph{PP-butterfly} consists of two triads  $ a $ and $ b, $ with
wave amplitudes
 $ B_{ja}, \ B_{jb}, $ $j=1,2,3$, and connecting mode, say $
 B_{1a}=B_{1b}= B_{1},$ is passive in both triads. The dynamical system
for PP-butterfly reads
 \bea \label{PP}
\begin{cases}
\dot{B}_{1}=  Z_a B_{2a}^*B_{3a} +   Z_b B_{2b}^*B_{3b}\,, \\
\dot{B}_{2a}=  Z_a B_{1}^* B_{3a}\,,\quad    \dot{B}_{2b}=  Z_b B_{1}^* B_{3b}\,, \\
\dot{B}_{3a}=  - Z_{a} B_{1} B_{2a} \,,\    \dot{B}_{3b}=  - Z_{b} B_{1} B_{2b}\ . \\
\end{cases}
 \eea%
 \begin{figure}[h]
\begin{center}\includegraphics[width=3cm,height=0.75cm]{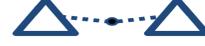}
\end{center}
\caption{\label{f:PP-but} NR-diagram for PP-butterfly.}
\end{figure}

We have studied this in \cite{BK09_1}; we present here the results in order to compare the dynamics of different butterfly
types and confirm the qualitative analysis given in \cite{KL-06}. Sys.(\ref{PP})
 has 3 quadratic CLs
analogous to (\ref{laws3waves}) and 1 cubic CL corresponding to its Hamiltonian:
 \bea \label{int-PP}
\begin{cases} I_{23a}=|B_{2a} |^2 + |B_{3a}|^2 \,, \quad
 I_{23b}= |B_{2b} |^2 + |B_{3b}|^2 \,, \\
 I_{ab}= |B_{1}|^2 + |B_{3a} |^2 + |B_{3b}|^2 \,, \\
H_{PP} = \operatorname{Im}(Z_a B_{1} B_{2a} B_{3a}^* + Z_b B_{1}
B_{2b} B_{3b}^*).
 \end{cases}
  \eea%

\emph{Standard  amplitude--phase  representation.} Here, one can rewrite the cubic conservation law as
 \be \label{m: Hamiltonian} H_{PP} = C_{PP} \left( Z_a C_{2a} C_{3a} \sin \varphi_a + Z_b
C_{2b} C_{3b} \sin \varphi_b\right).
 \ee
Here 
 \be \nonumber \label{ab-phases}
 \varphi_{a} = \theta_{1a} + \theta_{2a} -\theta_{3a}, \quad
\varphi_{b} = \theta_{1b} + \theta_{2b} -\theta_{3b} \ee
are dynamical phases and $C_{(PP)}$ is the real amplitude of a common mode in PP-butterfly. This allows us to reduce
Sys.(\ref{PP}) to only four real equations:
  \bea \label{PP-standart}
  \begin{cases}
\dot{C}_{3a} = - Z_a C_{PP}  C_{2a} \cos \varphi_a\,, \\
\dot{C}_{3b} =   - {Z_b} \, C_{PP}  C_{2b} \cos \varphi_b\,,   \\
\dot{\varphi}_a =  - H_{PP} (C_{PP}^{-2}+C_{2a}^{-2}-C_{3a}^{-2}),\\
\dot{\varphi}_b =  - H_{PP}
(C_{PP}^{-2}+C_{2b}^{-2}-C_{3b}^{-2}).
\end{cases}
 \eea
Now the overall dynamics of the PP-butterfly is confined to a $3$-dimensional manifold. The same can be done for the two
other types of butterflies.\\

\emph{Modified amplitude-phase representation.} The following change
of variables was suggested in \cite{BK09_1}:
 \be \nonumber \label{m: new coord} \alpha_a = \arctan\left(C_{3a}/C_{2a}\right),
 \ \
\alpha_b = \arctan\left(C_{3b}/C_{2b}\right)\,.
 \ee
with the inverse transformation being
 \be
 \begin{cases}
 C_{2a} = \sqrt{I_{23a}} \cos\alpha_a\,, \quad C_{3a} = \sqrt{I_{23a}}\sin\alpha_a\,,\\
 C_{2b} = \sqrt{I_{23b}} \cos\alpha_b\,, \quad C_{3b} = \sqrt{I_{23b}}\sin\alpha_b\,,
 \end{cases}
 \ee
This change of variables allows further substantial simplification
of (\ref{PP}) and (\ref{PP-standart}): %
\bea \label{m: alpha}
\begin{cases}
\dot{\alpha}_{a} = - Z_a C_{PP} \cos \varphi_a, \\
\dot{\alpha}_{b} = - Z_b C_{PP} \cos \varphi_b\,, \\
\dot{\varphi}_a  =   Z_a C_{PP} \left( \cot \alpha_a -
\tan \alpha_a\right)\sin \varphi_a - H_{PP}/C_{PP}^2 ,\\
\dot{\varphi}_b   =  {Z_b} C_{PP} \left( \cot \alpha_b -
\tan \alpha_b\right)\sin \varphi_b  - H_{PP}/C_{PP}^2.
\end{cases}
 \eea%
In these new variables, the amplitude $C_{PP}>0$ reads
 \be \label{m: amplitude C_1-PP} C_{PP} =
 \sqrt{I_{ab} - I_{23a} \sin^2 \alpha_a  - I_{23b} \sin^2 \alpha_b }\,
 \ee
and the Hamiltonian is now
 \bea \label{m:Hamiltonian_final}
 H_{PP} = \frac{C_{PP}} 2  Z_a I_{23a} \sin{\varphi_a} \sin{2\alpha_a}+ \nn \\
  \frac{C_{PP}} 2  Z_b I_{23b}  \sin{\varphi_b} \sin{2\alpha_b}.
 \eea
Eqs. (\ref{m: alpha})--(\ref{m:Hamiltonian_final}) represent the
final form of our three-dimensional general system in the modified
amplitude-phase presentation.

A few cases of integrability in quadratures of the PP-butterfly were
 presented in \cite{BK09_1}. The results are collected in Table
 \ref{t:ClustIntegr} below. Of course, the form of the conservation laws is arbitrary
 in the sense that any set of functionally independent CLs will be
 suitable. For instance, in the case $H_{PP} = 0$ we could choose the conserved quantity
 $$A_b = \sin{\varphi_b}
 \sin{2\alpha_b} \quad \mbox{instead of} \quad A_a = \sin{\varphi_a}
 \sin{2\alpha_a}$$
 (see Table \ref{t:ClustIntegr}, 1.2 PP) but not both because they are functionally
 dependent:
$$Z_a I_{23a} A_a + Z_b I_{23b} A_b \equiv 0.$$
Generally, we try to find the simplest presentation for our new constants of motion.

\subsection{Ray}
Analogously with the previous case, \emph{AA-butterfly} is a two-triad cluster with
a common mode which is A-mode in both triads,  $ B_{3a}=B_{3b}$. Dynamical system and Manley--Rowe constants read:%
\bea \label{AA}
\begin{cases}
\dot{B}_{1a}=  Z_{a} B_{2a}^* B_{3a}\,,\    \dot{B}_{1b}=  - Z_{b}  B_{2b}^*B_{3a}\,, \\
\dot{B}_{2a}=  Z_a B_{1a}^* B_{3a}\,,\quad    \dot{B}_{2b}=  Z_b B_{1b}^* B_{3a}\,, \\
\dot{B}_{3a}= - Z_a B_{1a}B_{2a} -   Z_b B_{1b}B_{2b} \ . \\
\end{cases}\\
\begin{cases}I_{12a}=|B_{1a} |^2 - |B_{2a}|^2 \,, \quad
 I_{12b}= |B_{1b} |^2 - |B_{2b}|^2 \,, \\ \label{AAint}
   I_{ab}=|B_{1a}|^2+ |B_{3a} |^2 + |B_{3b}|^2\ .
\end{cases}
 \eea%

\begin{figure}[h]
\begin{center}\includegraphics[width=3cm,height=0.75cm]{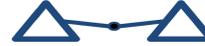}
\end{center}
\caption{\label{f:AA-but} NR-diagram for AA-butterfly.}
\end{figure}
The integrability of (\ref{AA}) can be investigated along the same lines as for (\ref{PP}) above. The analysis is omitted
here. We just partly outline one particular case of this cluster: $AA$-ray, which can be regarded as a degenerate
$AA$-butterfly, so that $\o_{1b} = \o_{2b} = \o_3/2$.

In this case, the dynamical system
obtained from first principles will have the form
\bea \label{AA-Ray}
\begin{cases}
\dot{B}_{1a} =  Z_{a} B_{2a}^* B_{3}\,,\    \dot{B}_{b}=   Z_{b}  B_{b}^*B_{3}\,, \\
\dot{B}_{2a} =  Z_a B_{1a}^* B_{3}\,,\quad \dot{B}_{3}= - Z_a
B_{1a}B_{2a} -  2 Z_b B_{b}^2 \ .
\end{cases}
 \eea%
Notice that there is a factor $2$ in the last term of last equation, which would not appear if we made the direct
substitution $B_{1b} = B_{2b} =B_b$ into system (\ref{AA}). Rather, the simple change of variables $B_{1b} = B_{2b} =
\sqrt{2} B_b$ will transform the AA-butterfly (\ref{AA}) into the ray equations (\ref{AA-Ray}). This means in particular
that integrable cases of AA-butterfly can be directly mapped to some integrable cases of AA-ray. Another interesting point
is that AA-ray cluster might also have a nice mechanical model - Wilberforce pendulum \cite{Ly09}, the problem is presently
under the study.

Conservation laws for AA-ray are inherited from conservation laws for
AA-butterfly:%
 \bea \label{CL ray}
\begin{cases}I_{12a}=|B_{1a} |^2 - |B_{2a}|^2 \,,\\
   I_{ab}=|B_{1a}|^2+ 2 |B_{b} |^2 + |B_{3}|^2\ \,,\\
   H_{\mathrm{ray}} = \operatorname{Im}(- Z_a B_{1a} B_{2a} B_{3}^* - 2 Z_b
   B_{b}^2 B_{3}^*),
\end{cases}
 \eea%
with dynamical phases 
 \be \nonumber \label{ab-phases-AA-Ray}
 \varphi_{a} = \theta_{1a} + \theta_{2a} -\theta_{3}, \quad
\varphi_{b} = 2 \theta_{b} -\theta_{3}. \ee
This reduces four complex equations (\ref{AA-Ray}) to only four real
ones:
  \bea\label{ray}
  \begin{cases}
\label{ray: C1a}  \dot{C}_{1a} =  Z_a   C_{2a} C_3 \cos \varphi_a\,, \\
\label{ray: Cb} \dot{C}_{b} =  {Z_b} C_{b} C_3 \cos \varphi_b\,,   \\
\label{ray: phi_a}\dot{\varphi}_a =  - Z_a C_3 \left(
\frac{C_{2a}}{C_{1a}}
+ \frac{C_{1a}}{C_{2a}}\right)\sin \varphi_a + H_{\mathrm{ray}}/C_3^2 \\
\label{ray: phi_b}\dot{\varphi}_b =  - 2 Z_b C_3 \sin
\varphi_b +H_{\mathrm{ray}}/C_3^2\,,
\end{cases}
 \eea
with Hamiltonian
 \be \label{ray: Hamiltonian} H_{\mathrm{ray}} =  - C_3 \left( Z_a C_{1a} C_{2a} \sin \varphi_a + 2 Z_b
C_{b}^2 \sin \varphi_b\right)
 \ee
in terms of the amplitudes and phases.

Consider the simple case when initially $\varphi_a = \varphi_b = 0$.
Then $ H_{\mathrm{ray}} =0$, phases remain zero for all times, and
the equations of motion reduce to
  \bea \label{ray0}
  \begin{cases}
\label{ray0: C1a}  \dot{C}_{1a}  = Z_a   C_{2a} C_3 \,, \\
\label{ray0: C2a}  \dot{C}_{2a}  =  Z_a   C_{1a} C_3 \,, \\
\label{ray0: Cb}   \dot{C}_{b}   =  {Z_b} C_{b} C_3 \,,\\
 \label{ray0: C3}  \dot{C}_{3}   =  - {Z_a} C_{1a} C_{2a} - 2 {Z_b} C_{b}^2 \,.
 \end{cases}
 \eea
with two Manley--Rowe constants of motion%
 \bea \label{CL ray0}
\begin{cases}I_{12a}=C_{1a}^2 - C_{2a}^2 \,,\\
   I_{ab}=C_{1a}^2+ 2 C_b^2 + C_3^2\ \,.
\end{cases}
 \eea%
and a new one
 \bea \label{NewCL ray0} H_{\mathrm{new}} =
2 Z_a \ln C_b + Z_b \ln \left(\frac{C_{2a} - C_{1a}}{C_{2a} +
C_{1a}}\right)\,,
 \eea%
where notations
 $$C_{1a} = \sqrt{I_{12a}} \cosh \a \,, \quad C_{2a}
= \sqrt{I_{12a}} \sinh \a $$
are used.

Notice that this way only non-polynomial additional conservation laws are obtained (see Table \ref{t:ClustIntegr}).

\begin{table*}
\begin{tabular}{||c|c|c|c||}
   \hline
      Conditions & & Additional CLs  \\
   \hline
\hline
 & & \\
      $\varphi_a = \varphi_b =  0$  & 1.1.& $Z_b
\arctan\left(\frac{C_{3a}}{C_{2a}}\right) - Z_a
\arctan\left(\frac{C_{3b}}{C_{2b}}\right)$  \\
 &PP-but. & \\
   \hline
      \hline
       & & \\
     $\varphi_a,\ \varphi_b \neq 0, \ H_{PP}=0$ & 1.2.& $A_a = \sin{\varphi_a} \sin{2\alpha_a},$ \quad with  $\alpha_a=\arctan\left(C_{3a}/C_{2a}\right)$  \\
 &PP-but. & \\
         \hline
          & & \\
      & 1.3.&   $(1+\frac{Z_b}{Z_a}) \arccos\left(\frac{\cos
2 \alpha_a}{\sqrt{1-A_a^2}}\right) - (1+\frac{Z_b}{Z_a})
\arccos\left(\frac{\cos 2
\alpha_b}{\sqrt{1-A_b^2}}\right) $   \\
 &PP-but. & \\
 & & with $\alpha_b = \arctan\left(C_{3b}/C_{2b}\right)$ and $A_b = \sin{\varphi_b}
 \sin{2\alpha_b}$ \\
  & & \\
   \hline
      \hline
       & & \\
      $H_{PP} \neq 0, \ Z_a=Z_b$  & 1.4. & $ C_{2a}^2 C_{3a}^2 + C_{2b}^2 C_{3b}^2 +
 2 C_{2a} C_{3a} C_{2b} C_{3b} \cos(\varphi_a-\varphi_b)- C_1^2(C_1^2 - C_{2a}^2+ C_{3a}^2 - C_{2b}^2 + C_{3b}^2)$\\
   &PP-but. & \\
\hline
\hline
        & & \\
        &2.1. & $2 Z_a \ln C_b + Z_b \ln \Big[(C_{2a} - C_{1a})/(C_{2a} +
C_{1a})\Big]$\\
 $\varphi_a = \varphi_b =  0$        &AA-ray & with  $C_{1a} = \sqrt{I_{12a}} \cosh \a \,, \quad C_{2a}
= \sqrt{I_{12a}} \sinh \a \,, \quad I_{12a}=C_{1a}^2 - C_{2a}^2 $\\
        & & \\
\hline
\hline
 \end{tabular}
\caption{Examples of non-polynomial conservation laws. }\label{t:ClustIntegr}
\end{table*}

\subsection{Star}
\begin{figure}[h]
\begin{center}
\includegraphics[width=3cm,height=1.5cm]{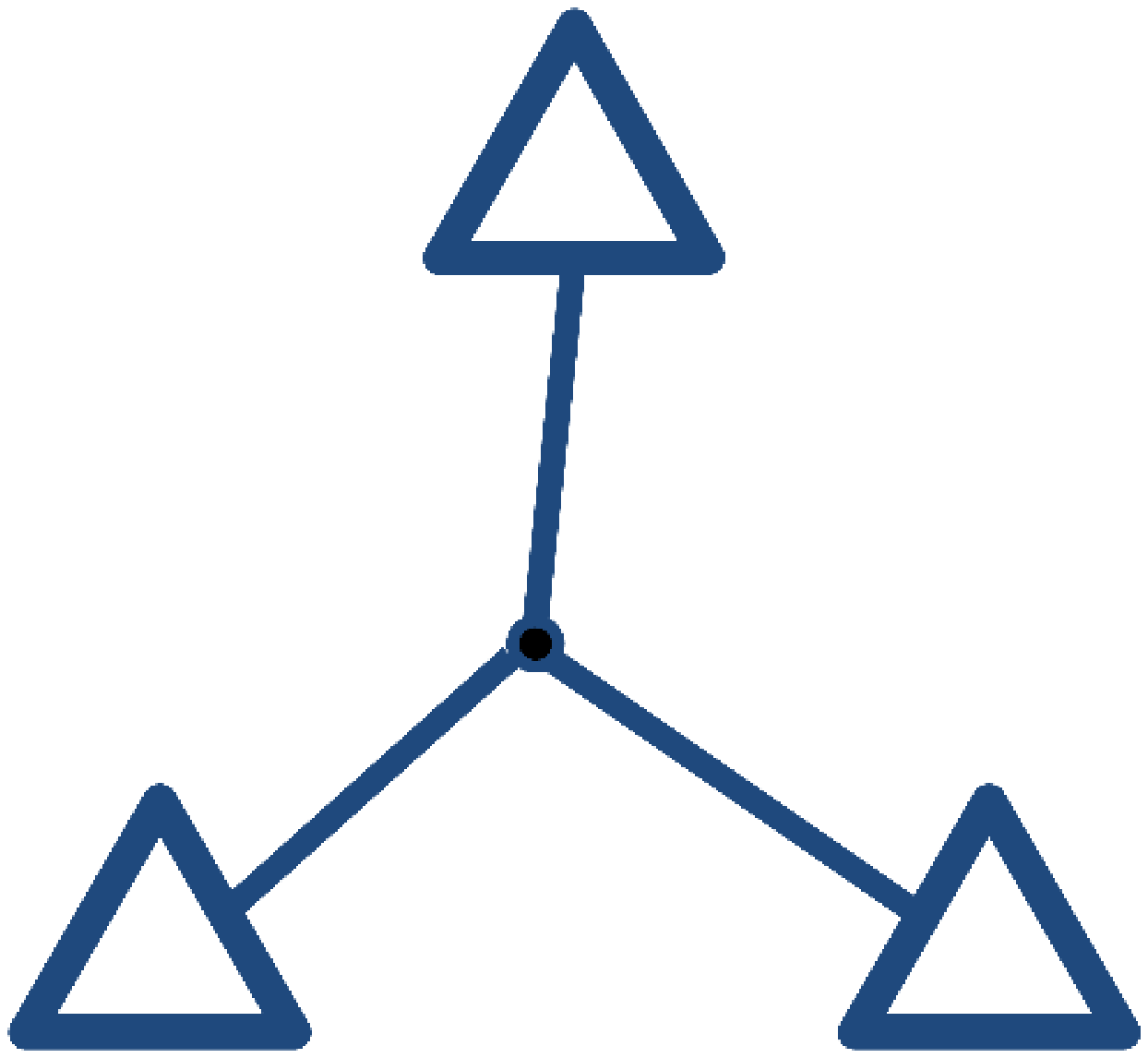}
\includegraphics[width=3cm,height=1.5cm]{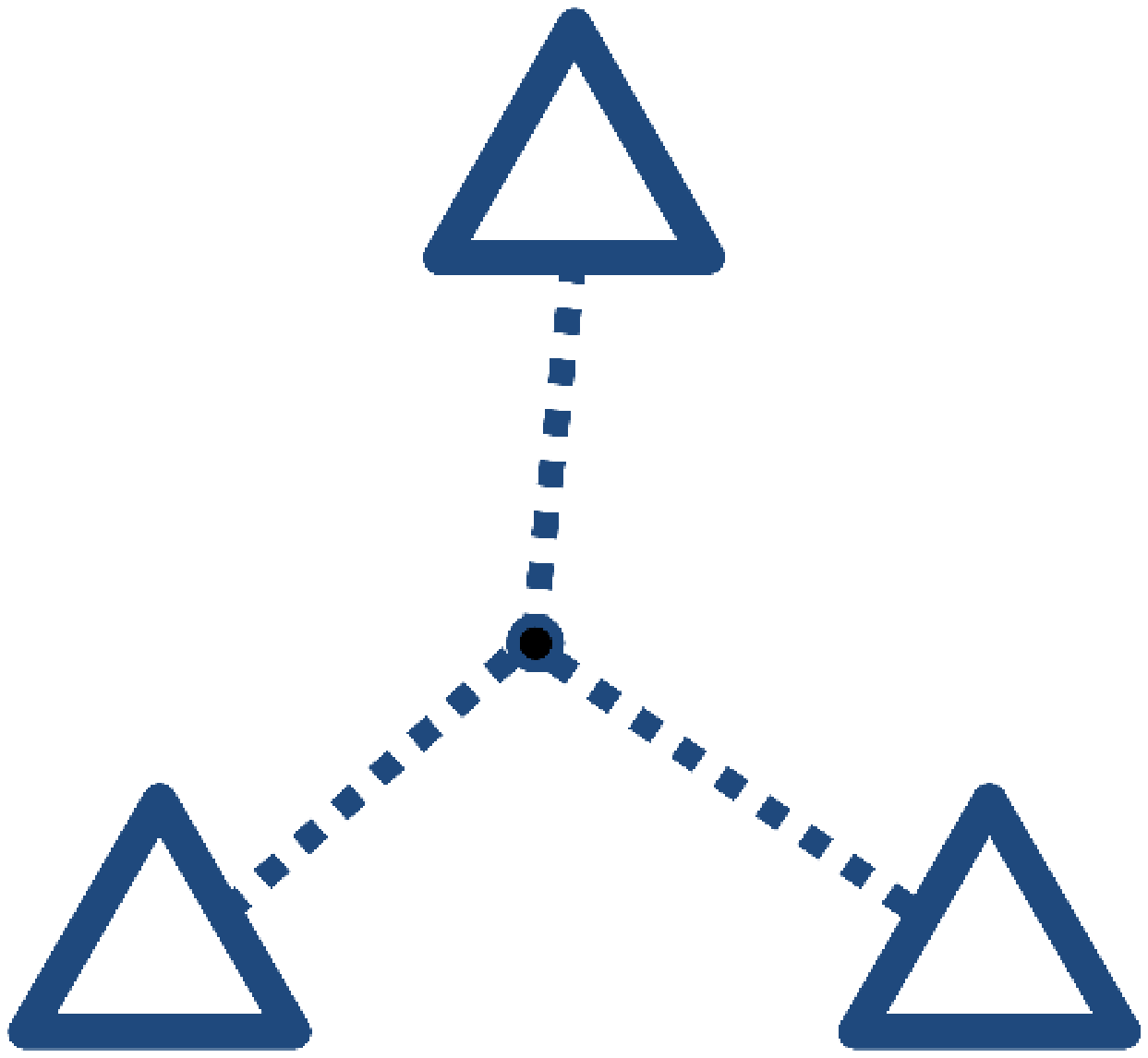}
\includegraphics[width=3cm,height=1.5cm]{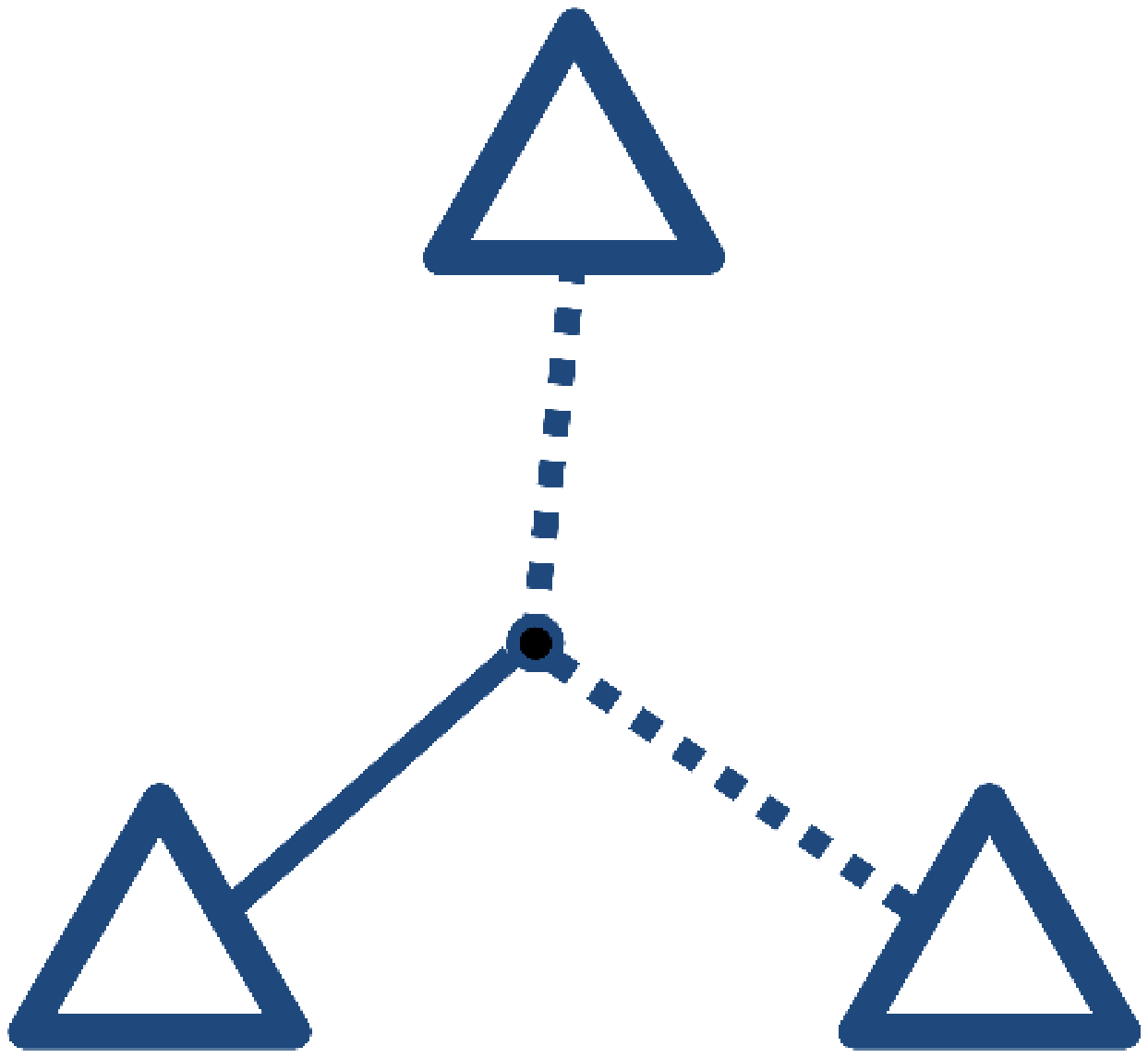}
\includegraphics[width=3cm,height=1.5cm]{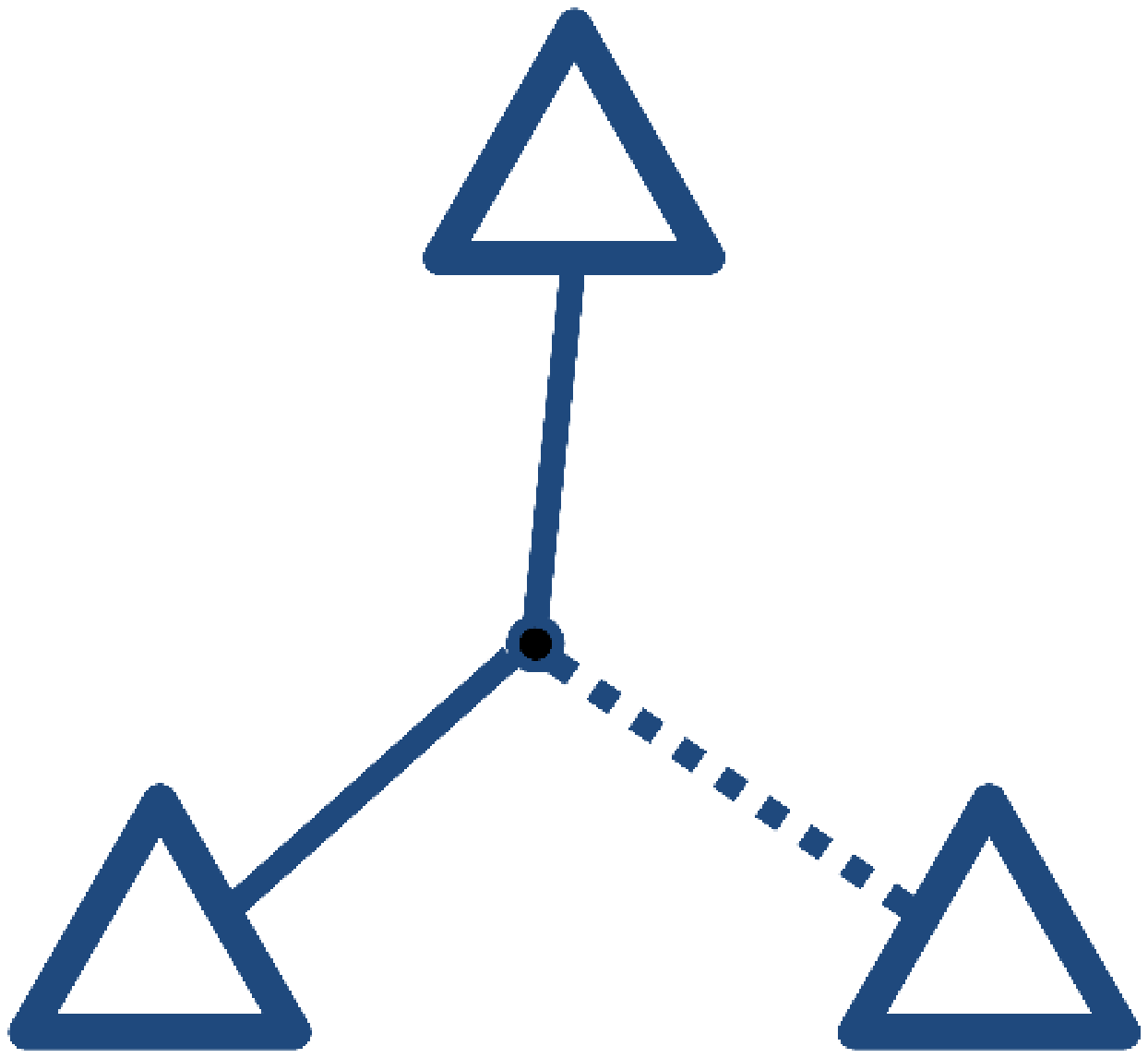}
\end{center}
\caption{\label{f:stars} NR-diagrams for 3-star clusters. From right to left, from up to down: 3-star-A (three
A-connections), 3-star-P (three P-connections), 3-star-1A-2P (one A-connection and two P-connections), 3-star-2A-1P (two
A-connections and one P-connection)}
\end{figure}

A cluster of \emph{N} triads, all connected \emph{via} one common mode is called \emph{N}-star cluster. Again,
integrability of \emph{N}-star depends on the types these connecting modes have in each triad of a cluster. NR-diagrams for
all possible types of 3-stars are shown in Fig. \ref{f:stars}.

\emph{N}-star cluster is probably the only presently known type of cluster for which an analytical study has been performed
for \emph{arbitrary} finite number $N$.  The main idea can be briefly formulated as follows. \emph{N}-star cluster has
\emph{2N+1} degrees of freedom, \emph{N+1} Manley--Rowe constants of motion and one Hamiltonian, that is, we already have
\emph{N+2} independent first integrals in involution. To find $N-1$ additional integrals of motion, one can use
construction of Lax operators, Painlev\'{e} analysis and irreducible forms (see \cite{MCL83a,MCL83b,Ver68a,Ver68b}, etc.;
terminology used therein is pump and daughter wave for A- and P-mode correspondingly). The dynamical system, say for
\emph{N}-star-A, is regarded in the form
 \bea \label{Nstar-A} \dot{B}_{1j}=i \, \lambda_j B_3 \dot{B}_{2j}, \quad
\dot{B}_{2j}=i \, \lambda_j B_3 \dot{B}_{1j},\\
\dot{B}_{3} = i \, \sum_{j=1}^{N} \lambda_j B_{1j}B_{2j}. \eea
 Additional conservation laws found this way have necessarily
\emph{polynomial form}.  The  results for a
generic N-star cluster are as follows: N-star-A (with all A-connections) and  N-star-P (with all P-connections) are
integrable for \emph{arbitrary initial conditions}  if
$$\lambda_j=\frac{1}{2} \quad \mbox{or} \quad 1 \quad \mbox{or} \quad 2, $$
examples of corresponding NR-diagrams shown in the Fig.\ref{f:stars}, upper panel, for $N=3$. $N$-star cluster with mixed
A- and P-connections (Fig.\ref{f:stars}, lower panel) has no additional polynomial conservation laws. Complete set of
additional polynomial conservation laws for integrable N-star cluster is omitted here for sake of place, and it can be found in
\cite{Ver68b}. Example for the case of AA-butterfly with  $Z_a =2 \,Z_b$ reads
\bea 4(B_1B_2B_4^*B_5^*+B_1^*B_2^*B_4B_5)(B_1B_1^*+B_2B_2^*)\nonumber\\ -2(B_3B_1^*B_2^*+B_3^*B_1B_2)^2-[(B_1B_1^*+B_2B_2^*)^2 \nonumber\\
+4B_1B_1^*B_2B_2^*](B_4B_4^*+B_5B_5^*). \eea

The general way to investigate integrability of a generic cluster would be to apply the theory of normal forms (see, for
instance, \cite{Mu03,Na93}) for each dynamical system describing a resonance cluster is a normal form. But of course, most
generic clusters demonstrate chaotic behavior and numerical investigations are unavoidable.

\begin{figure}[h]
\begin{center}
\includegraphics[width=9cm,height=6cm]{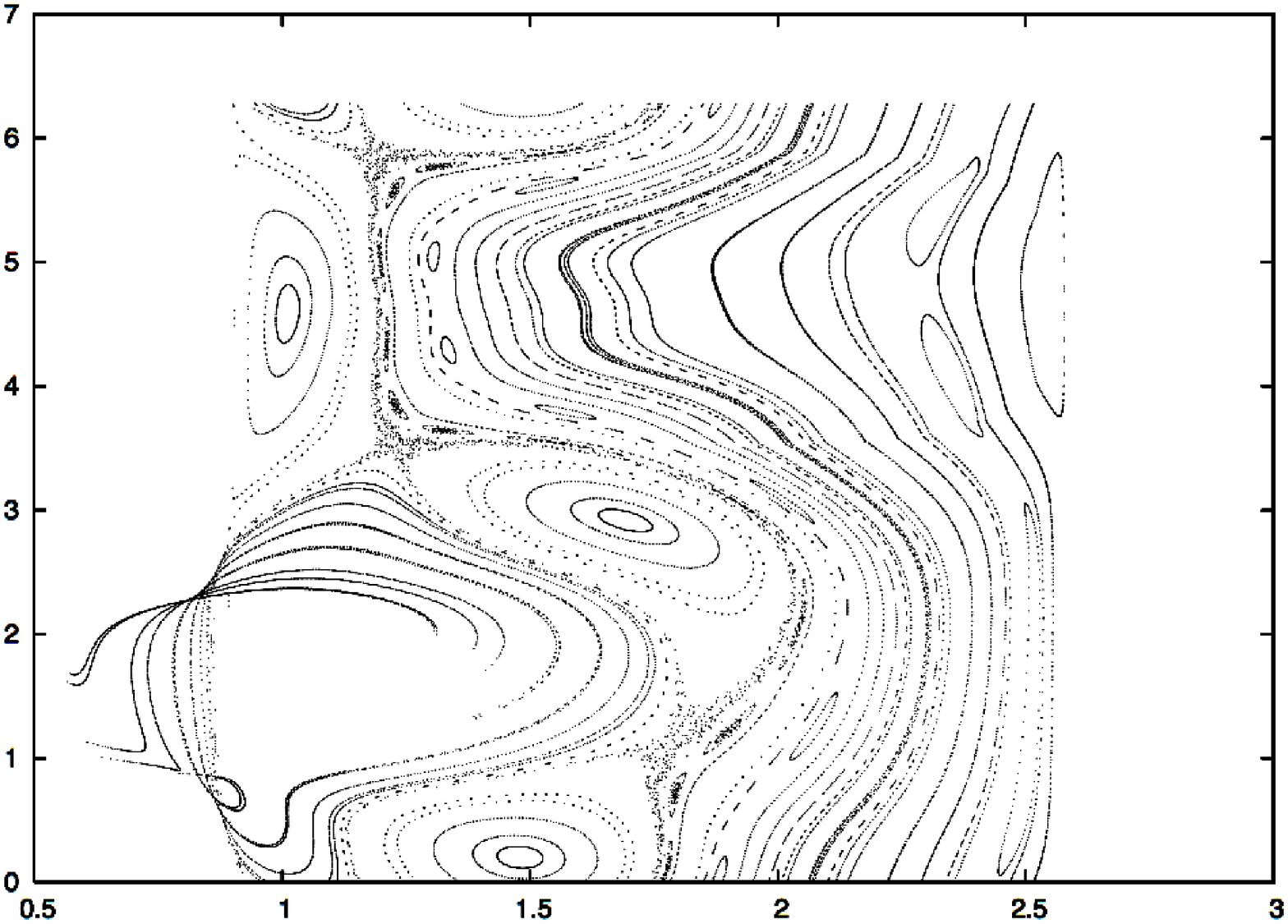}
\end{center}
\caption{\label{f:chaos} Example of Poincar\'e section for PP-butterfly with ${Z_a}/{Z_b}=3/4.$}
\end{figure}

Here we outline briefly which facts are important in order to perform sensible numerical simulations with resonance
clusters.
The fact that our systems are Hamiltonian, allows us
to perform numerical simulations based on the Hamiltonian expansion of the corresponding dynamical system. Poincar\'e
sections are well-known instruments to show clearly whether or not a dynamical system demonstrates chaotic behavior, so all
the results of our numerical simulations are illustrated by the corresponding Poincar\'e maps.
One of the most tedious and time-consuming parts of the corresponding simulations is the choice of initial conditions. A special
procedure has been worked out \cite{Le08}  that guarantees a uniform distribution of initial conditions according to Liouville measure,
and assures as well that all conservation laws have the same value on each Poincar\'e section.
 In Fig.\ref{f:chaos} an example of Poincar\'e section is shown.
To trace effects due to the dynamical phases one has to use amplitude-phase representations, of course.

\section{Coupling coefficient}
As we have shown above, the integrability of a resonance cluster depends on the magnitude of the corresponding coupling
coefficients. Expressions for  coupling coefficients in \emph{canonical
variables} has been deduced for various types of wave systems possessing three-wave resonances: irrotational capillary waves \cite{zak2}; rotational capillary waves \cite{CKW09}; irrotational gravity-capillary waves \cite{McG65}; rotational gravity-capillary waves \cite{Jo97}, drift waves \cite{Pit98}, etc. They usually have a nice compact form, for instance, coupling coefficient $V^3_{12}$ for irrotational gravity-capillary water waves reads
$$ \frac{(\o_2^2-\o_2 \o_3 + \o_3^2)}{\o_1}k_1 -\o_2 k_2  + \o_3 k_3, $$
where $\o=(g \, k + \sigma k^3)^{1/2}$ and $g$ and $\sigma$ are gravity acceleration and surface tension correspondingly.
However,  transformation of these expressions from the canonical to
\emph{physical variables}  is not an easy task.

On the other hand, the application of multi-scale methods yields expressions for the coupling coefficients directly in
physical variables. For instance, coupling coefficients of the system of three resonantly interacting atmospheric planetary
waves, with $\o \sim m/[n(n+1)]$, have the form \cite{all08}
 \bea \label{ch1:Z-coef}
\begin{cases}
Z[n_2(n_2+1)-n_3(n_3+1)]/[n_1(n_1+1)], \\
Z[n_3(n_3+1)-n_1(n_1+1)]/[n_2(n_2+1)], \\
Z[n_2(n_2+1)-n_1(n_1+1)]/[n_3(n_3+1)],
\end{cases}
\eea
with
\be \label{ch1:Z}
Z= \int_{-\pi/2}^{\pi/2}[m_2P^{(2)}\frac{d}{d \varphi}P^{(1)}-
m_1P^{(1)}\frac{d}{d \varphi}P^{(2)}] \frac{d}{d \varphi}P^{(3)} d \varphi. \nonumber \ee
 Here two spherical space variables are
 the
latitude $\varphi$,  $-\pi/2 \leq \varphi \leq \pi/2$, and the longitude $\lambda$, $0 \leq \lambda \leq 2\pi$, and the
notation $P^{(j)}$ is used for $ P_{n_j}^{m_j} (\sin \ph) $ which is the associated Legendre function of degree $n_j$ and
order $m_j$.

The multi-scale method is quite straightforward and can be programmed in some symbolical language \cite{all08}. However, only
numerical magnitudes of the coupling coefficients have been computed for selected solutions of
the resonance conditions, not an explicit algebraic formulas.
 The problem is due to some "bags" in \emph{Mathematica} in computing integrals of the form
 \be
 \label{eq:search}
 \int_0^{2\pi}\sin(mx)\sin(nx)dx, \quad \mbox{with} \quad
m,n\in\mathbb{N},\ee
more discussion can be found in \cite{all08}.

For completeness of presentation we show below a part of the formula for one coupling coefficient, obtained by a
combination of symbolical programming in \emph{Mathematica} and automatic search-and-replace of formulas of the type
(\ref{eq:search}) -- which \emph{Mathematica} cannot handle directly -- for ocean planetary waves with vanishing boundary
conditions, $\o \sim (m^2+n^2)^{-1/2}$, in physical variables. The general form of the coupling coefficients is
\be \label{appen}
\a_j/[(m_j^2 + n_j^2)\pi^2],\quad j=1,2,3,
\ee
 with coefficients $\a_j $ being functions of the wavenumbers $m_1,\, n_1, \, \cdots, m_3,\, n_3.$
Notations \textbf{E} and \textbf{I} below are for exponent $\exp$ and imaginary unit $i$, all other notations are
self-explanatory. Notation (...) at the end is used for eight more pages, needed to accomplish this formula.

$\a_1=\frac {1} {16} \times$
 \begin{widetext}
\begin{verbatim}
(E^(-I (m1 + m2 + m3 + Sqrt[m1^2 + n1^2] + Sqrt[m2^2 + n2^2] - Sqrt[m3^2 +n3^2])\[Pi])
((I (m2^2 m3 n2 - 2 m3^3 n2 + m3 n2^3 + m2 m3 n2 Sqrt[m2^2 + n2^2] + 2 m2^3 n3 - m2 m3^2 n3
+ 2 m2 n2^2 n3 + 2 m2^2 Sqrt[m2^2 + n2^2] n3 - m3^2 Sqrt[m2^2 + n2^2] n3 + n2^2 Sqrt[m2^2
+ n2^2] n3 - 2 m3 n2 n3^2 - m2 n3^3 - Sqrt[m2^2 + n2^2] n3^3 + m2^2 n2 Sqrt[m3^2 + n3^2]
- 2 m3^2 n2 Sqrt[m3^2 + n3^2] + n2^3 Sqrt[m3^2 + n3^2] + m2 n2 Sqrt[m2^2 + n2^2] Sqrt[m3^2
+ n3^2] - m2 m3 n3 Sqrt[m3^2 + n3^2] - m3 Sqrt[m2^2 + n2^2] n3 Sqrt[m3^2 + n3^2]
- n2 n3^2 Sqrt[m3^2 + n3^2]))/(m1 + m2 - m3 + Sqrt[m1^2 + n1^2] + Sqrt[m2^2 + n2^2]
- Sqrt[m3^2 + n3^2]) + (I (m2^2 m3 n2 - 2 m3^3 n2 + m3 n2^3 + m2 m3 n2 Sqrt[m2^2 + n2^2]
+ 2 m2^3 n3 - m2 m3^2 n3 + 2 m2 n2^2 n3 + 2 m2^2 Sqrt[m2^2 + n2^2] n3 - m3^2 Sqrt[m2^2
+ n2^2] n3 + n2^2 Sqrt[m2^2 + n2^2] n3 - 2 m3 n2 n3^2 - m2 n3^3 - Sqrt[m2^2 + n2^2] n3^3
+ m2^2 n2 Sqrt[m3^2 + n3^2] - 2 m3^2 n2 Sqrt[m3^2 + n3^2] + n2^3 Sqrt[m3^2 + n3^2]
+ m2 n2 Sqrt[m2^2 + n2^2] Sqrt[m3^2 + n3^2] - m2 m3 n3 Sqrt[m3^2 + n3^2] - m3 Sqrt[m2^2
+ n2^2] n3 Sqrt[m3^2 + n3^2] - n2 n3^2 Sqrt[m3^2 + n3^2]))/(m1 - m2 + m3 - Sqrt[m1^2
+ n1^2] - Sqrt[m2^2 + n2^2] + Sqrt[m3^2 + n3^2]) - (I (m2^2 m3 n2 - 2 m3^3 n2 + m3 n2^3
- (...)
\end{verbatim}
 \end{widetext}

\section{Summary}
(..to be added:  numerical simulations, NR-reduced models..)\\

\noindent {\bf Acknowledgements}. The authors are very much obliged
to the organizing committee of the workshop ``INTEGRABLE SYSTEMS AND
THE TRANSITION TO CHAOS II," who provided an excellent opportunity
to meet and work together. They also highly appreciate the
hospitality of Centro Internacional de Ciencias (Cuernavaca,
Mexico), where part of the work was accomplished. Authors would like
to thank A. Degasperis, S. Lombardi and F. Leyvraz for interesting
and fruitful discussions.  M.B. acknowledge the support of the
 Transnational Access Programme
at RISC-Linz, funded by European Commission Framework 6 Programme
for Integrated Infrastructures Initiatives under the project SCIEnce
(Contract No. 026133). E.K. acknowledges the support of the Austrian
Science Foundation (FWF) under project
 P20164-N18 ``Discrete resonances in nonlinear wave systems".

\end{document}